\documentclass[11pt,a4paper]{article}
\usepackage{jheppub}
\usepackage{slashed}
\usepackage{cancel}
\usepackage{xcolor}

\makeatletter
\def\@fpheader{\relax}
\makeatother

\usepackage{ulem}

\usepackage[czech,english]{babel}
\usepackage{graphicx}
\usepackage{yfonts}
\usepackage{amsmath,amsfonts,amssymb}
\usepackage{url}
\usepackage{mathabx}
\usepackage{ulem}
\usepackage{mathtools}

\DeclareMathOperator{\MyProd}{\scalebox{1.4}{$\mathrm{I\kern-0.2ex I}$}}

\preprint{LCTP-23-06}

\title{Universal Cardy-Like Behavior of 3D Partition Functions from Supersymmetric Localization}

\author[a]{Antonio Amariti,}

\emailAdd{antonio.amariti@mi.infn.it}

\author[b]{ Jun Nian,}

\emailAdd{nianjun@ucas.ac.cn}

\author[c, d, e]{ Leopoldo A. Pando Zayas,}

\emailAdd{lpandoz@umich.edu}

\author[a,f]{ Alessia Segati }

\emailAdd{alessia.segati@mi.infn.it}

\affiliation[a]{INFN, Sezione di Milano, 20133 Milano, Italy}
\affiliation[b]{International Centre for Theoretical Physics Asia-Pacific, University of Chinese Academy of\\ Sciences, 100190 Beijing, China}
\affiliation[c]{Leinweber Center for Theoretical Physics, University of Michigan, Ann Arbor, MI 48109, USA}
\affiliation[d]{The Abdus Salam International Centre for Theoretical Physics, 34014 Trieste, Italy}
\affiliation[e]{School of Natural Sciences, Institute for Advanced Study,  Princeton,  NJ 08540, USA}
\affiliation[f]{Dipartimento di Fisica, Universit\`a degli Studi di Milano, Via Celoria 16, I-20133 Milano,
Italy}

\abstract{We investigate 3d $\mathcal{N}=2$ supersymmetric gauge theories on $S^1 \times S^2$ and the corresponding 2d effective field theories arising in the limit of small ratio of radii,  $\beta=R_{S^1}/R_{S^2}\to 0$.  We evaluate the exact partition function of these theories in the framework of supersymmetric localization on curved backgrounds.  As a result, we establish a finite-$N$ map between a particular,  superconformal-index-inspired partition function and the topologically twisted index.  Taking the large-$N$ limit of the partition functions,  we reproduce the entropy functions of either spherically symmetric, magnetically charged, or rotating, electrically charged asymptotically AdS$_4$ black holes.  We then recast the problem of evaluating the 3d partition functions directly in the framework of rigid supersymmetry.  By carefully tracking the background fields,  we find that in the small-$\beta$ limit, the partition functions of these 3d large-$N$ superconformal field theories have a universal behavior related to the coefficients of the R-symmetry or flavor symmetry 2-point current correlation functions,  thus obtaining a universal Cardy-like formula for 3d $\mathcal{N}=2$ superconformal field theories.}


\arxivnumber{}

\newcommand{\bea}{\begin{eqnarray}}
\newcommand{\eea}{\end{eqnarray}}

\newcommand{\be}{\begin{equation}}
\newcommand{\ee}{\end{equation}}

\begin{document}

\maketitle

\section{Introduction}

A microscopic foundation for the entropy of large classes of rotating, electrically charged asymptotically AdS black holes has recently been provided via supersymmetric partition functions on the field theory side of the AdS/CFT correspondence  \cite{Cabo-Bizet:2018ehj, Choi:2018hmj, Benini:2018ywd, Choi:2019miv, Kantor:2019lfo, Nahmgoong:2019hko, Choi:2019zpz, Nian:2019pxj}.  These supersymmetric partition functions have the remarkable property that they take particularly simple forms in a so-called Cardy-like limit.  Recall that in two dimensions, the Cardy formula states that $\log\, Z_{S^1_\beta\times S^1}=\frac{\pi^2}{3}\frac{c}{\beta}+\ldots$,  thus determining the partition function on a torus in terms of the central charge, $c$, and the size of a shrinking circle, $\beta \to 0$,  \cite{Cardy:1986ie}.  For higher dimensional theories, defined on $S^1\times S^d$, the corresponding Cardy-like limit involves an expansion in vanishing $R_{S^1}/R_{S^d}\to 0$.  One of our goals in this manuscript is to explore such limit for 3d ${\cal N}=2$  superconformal field theories (SCFT) with the aim of establishing an analogous Cardy-like formula.

One can naturally view the process of placing a ($d+1$)-dimensional superconformal field theory on $S^1\times S^{d}$ while shrinking $S^1$ through the natural lens of a $d$-dimensional effective field theory (EFT). This is indeed a fruitful path to take and has led to various Cardy-like formulas for theories in $d+1=4,6$ \cite{DiPietro:2014bca,  ArabiArdehali:2015iow,  Chang:2019uag}.  These impressive Cardy-like results, however, require significant modifications to treat the case relevant for rotating, electrically charged AdS black holes where the $S^1$ circle is fibered over $S^{d}$,  as properly described in the framework of a Cardy-like limit in \cite{Choi:2018hmj,  Honda:2019cio,  ArabiArdehali:2019tdm,  Kim:2019yrz,  Nahmgoong:2019hko,  Choi:2019zpz}.  A clarifying step in the direction of better understanding the Cardy-like limit was taken in \cite{Cassani:2021fyv,  ArabiArdehali:2021nsx}, who emphasized the analytic structure of the ${\cal N}=1$ superconformal index to re-derive the four-dimensional  Cardy-like formula on the second sheet from the EFT point of view.  An essential ingredient in the EFT approach - the presence of certain Chern-Simons theory - was anticipated by direct computations of the superconformal index (SCI) in \cite{GonzalezLezcano:2020yeb,  Amariti:2020jyx,  Amariti:2021ubd}.

Largely inspired by the 3d EFT treatment of the SCI of 4d ${\cal N}=4$ SYM \cite{Cassani:2021fyv,  ArabiArdehali:2021nsx},  and emboldened by a plethora of direct computations \cite{Gang:2019uay,Bobev:2019zmz,  Benini:2019dyp, Choi:2019dfu,  Hosseini:2022vho,  GonzalezLezcano:2022hcf,  Bobev:2022wem,  Bobev:2023lkx, Bobev:2024mqw},  we discuss in this paper the 2d effective field theories obtained by the Kaluza-Klein dimensional reduction of the 3d SCFTs on $S_\beta^1 \times S^2$. We consider a Cardy limit of the form:
\be\label{eq:CardyLikeLimit}
  \beta \to 0\, ,\quad \text{for the non-rotating case};\quad \omega \sim \beta \to 0\, ,\quad \text{for the rotating case},
\ee
where $\omega \equiv 2\, \Omega\, \beta$ with $\Omega$ denoting the angular velocity of $S^2$.

Our analysis exploits two main tools: 1) A two-dimensional perspective for localization of 3d theories; 2) The equivalence, under rigid supersymmetry, of certain partition functions. The prototype 3d partition functions we examine can ultimately be viewed as natural generalizations of the 2d A-twist introduced by Witten \cite{Witten:1988ze}. For example, we will utilize and generalize the systematic discussion of some of the partition functions presented in \cite{Closset:2017zgf}. In particular, we are going to use the 2d perspective as the answer for the 2d EFT description of 3d ${\cal N}=2$  field theories \cite{Closset:2017zgf,  Closset:2019hyt} (see \cite{Hwang:2018riu} for an analogous 4d to 2d treatment).  In this paper,  we compute the supersymmetric partition functions on $S_\beta^1 \times S^2$ for several explicit examples of supersymmetric field theories on various backgrounds. Namely, we present results for the following four theories: (i)  the ABJM theory,  (ii) the mass deformed ABJM theory, (iii) the Chern-Simons theory with three adjoint   chiral multiplets whose gravity dual is the AdS$_4$ black hole in massive IIA theory,  and (iv) generalizations of quiver Chern-Simons matter theories with $\sim N^{5/3}$ scaling. In the large-$N$ limit, these 3d supersymmetric partition functions, when defined in the appropriate backgrounds, reproduce the entropy functions of the corresponding AdS$_4$ black holes,  either rotating electrically charged or spherically symmetric magnetically charged ones. As anticipated by Benini, Hristov, and Zaffaroni \cite{Benini:2015eyy}, in the spherically symmetric magnetically charged asymptotically AdS$_4$ black holes, we observe that the corresponding 3d supersymmetric partition function subjected to a judicious choice of background fields coincides with the topologically twisted index (TTI).

The rigid supersymmetry framework allows the comparison of partition functions computed in conceptually different backgrounds. We find, for example, that the partition function of a theory defined on (rotating) $S_\beta^1 \times S^2$ can be related to the topologically twisted theory on $S^1 \times S^2$ via the following identification:
\be\label{eq:Map}
  Z_{S_\beta^1 \times S^2} (r_i,\, \mathfrak{n}_i = 0) = \mathcal{I}_{TTI} \left(\mathfrak{n}_i = \frac{r_i}{2} \right)\, ,
\ee
where $r_i$ and $\mathfrak{n}_i$ denote the R-charges and the magnetic fluxes of the $i$-th flavor; the above identification is valid at finite $N$.  Implementing a few extra assumptions on the left-hand side of \eqref{eq:Map}, we arrive at an expression similar to the superconformal index.  A similar relation between the SCI and TTI has been noticed before in the literature by explicit computations. The authors of \cite{Choi:2019dfu} obtained it in the large-$N$ limit and,  more recently, a high-precision numerical analysis in the Cardy-like limit established the relation for finite $N$ \cite{Bobev:2022wem, Bobev:2024mqw}.  Here,  rather than through direct computations,  we arrive at the above relation by carefully analyzing the supersymmetric backgrounds on which the respective theories are defined within the framework of rigid supersymmetry.

In the context of rigid supersymmetry \cite{Festuccia:2011ws}, one defines a supersymmetric field theory on a curved space via its coupling to a supergravity multiplet which also determines the background fields. From this point of view, one can also leverage the properties of certain correlators into a concrete expansion of the appropriate partition function. For example, we slightly generalize results presented in \cite{Barnes:2005bw,  Closset:2012vg} to show that in the above Cardy-like limit,  \eqref{eq:CardyLikeLimit}, the free energies of 3d SCFTs defined on $S_\beta^1 \times S^2$ have a universal behavior related to the coefficients of the R-symmetry or the flavor symmetry 2-point correlation functions:
\be
\label{Cardy-like_Formula}
  \textrm{Re}\left( F_{S^1 \times S^2} \right) = - \frac{1}{2} (m \overline{L})^2\, \tau_{aa} + \mathcal{O} (\beta)\, ,\quad \textrm{Im}\left( F_{S^1 \times S^2} \right) \,\,\propto\,\, - \frac{1}{2} (m \overline{L})^2\, \kappa_{aa}\, ,
\ee
which we call the 3d Cardy-like formula.  Here,  $m$ and $\overline{L}$ denote the background value of the scalar in the vector multiplet and the characteristic length scale of the 3d curved background, respectively. Depending on whether $S^2$ is rotating or not,  $\overline{L}$ can be $\beta$ or $\bar{L} = \beta - 2 \pi i/ \Omega$ with the angular velocity $\Omega$.  Moreover,  $\tau_{aa}$ is the coefficient of the current-current two-point correlation function,  which can be either $\tau_{ff}$ or $\tau_{RR}$,  depending on whether the 3d SCFT has extra flavor symmetries. The term $\kappa_{aa}$ describes the contact terms in the correlator of two currents.

When examining Cardy-like formulas, there is a known asymmetry between even and odd dimensions. For even-dimensional field theories, one naturally expects terms containing anomalies. For odd-dimensional field theories, a natural candidate has been the free energy on the corresponding sphere, $F_{S^{d}}$ for $d=3,5$ \cite{Hosseini:2016tor,  Bobev:2019zmz,  Choi:2019dfu,  Crichigno:2020ouj}. Our result (\ref{Cardy-like_Formula}) clarifies that the proper object is the coefficient in the correlation function of two currents, $\tau_{RR}$. In the particular example of 3d SCFTs scaling as $\sim N^{5/3}$, we found a relation among the 2d central charge $c_r$,  the 4d anomaly coefficient $a$,  and the 3d coefficient $\tau_{RR}$,  consistent with the known extremization principles in the literature.  We plan to expand on these cross-dimensional connections in a separate work.

This paper is organized as follows.  In Sec.~\ref{sec:Bgd}, we review the supersymmetric curved backgrounds relevant to the discussions in this paper.  In Sec.~\ref{sec:2D}, we compute the supersymmetric partition functions for various examples of 3d $\mathcal{N}=2$ SCFTs on $S_\beta^1 \times S^2$ and take the large-$N$ limit.  We also discuss the corresponding 2d effective field theories in the Cardy-like limit \eqref{eq:CardyLikeLimit}.  In Sec.~\ref{sec:3D}, the relation between the free energy and the coefficients of the current-current correlation functions on $S_\beta^1 \times S^2$ will be discussed,  and we will revisit the same examples from the 3d perspective.  Based on the universal behaviors of the free energy of these examples in the Cardy-like limit,  we propose a Cardy-like formula for 3d large-$N$ $\mathcal{N}=2$ SCFTs.  Some discussions and future directions are presented in Sec.~\ref{sec:Discussion}.

\section{Review of 3d Rigid Supersymmetric Backgrounds}\label{sec:Bgd}

We are interested in supersymmetric theories on the following background:
\begin{align}
  ds^2 & = - dt^2 + L^2 \Big[d\theta^2 + \textrm{sin}^2 \theta \, (d\varphi + \Omega dt)^2\Big] \nonumber\\
  {} & = d\tau^2 + L^2 \Big[d\theta^2 + \textrm{sin}^2 \theta \, (d\varphi - i \Omega d\tau)^2\Big]\, ,\label{eq:metric}
\end{align}
which corresponds to the asymptotic metric on the boundary of a rotating, electrically charged  AdS$_4$ black hole. There are two equivalent perspectives that can be taken regarding the above metric. The first one views the passage from the first line to the second as a mere change of coordinates $t=-i\tau$.  In this context,  $\tau$ is purely imaginary,  and $\Omega$ is a real constant,  such that the whole metric \eqref{eq:metric} remains real. One can also/further consider a Wick rotation  which rotates $\tau$ from the imaginary axis to the real axis in the complex plane; in this latter situation one should correspondingly rotate the parameter $\Omega$ from a real constant to a purely imaginary constant. In this paper,  we work with real-valued $\tau$,  while $\Omega$ can be either real or purely imaginary,  which will be mentioned explicitly in the context.

In the metric \eqref{eq:metric},  the $S^2$ has the radius $L$,  and the $S^1$ has a real period
\be\label{eq:OldPeriod}
  \tau \sim \tau + \beta\, ,
\ee
where the period $\beta$ is independent of $L$.  As discussed in \cite{Nian:2019pxj},  the metric \eqref{eq:metric} can be written into a standard transversely holomorphic foliation (THF) form (see also the very pedagogical review \cite{Closset:2019hyt}):
\be\label{eq:metric-2}
  ds^2 = f^2 (d\chi + a dz + \bar{a} d\bar{z})^2 + c^2 dz d\bar{z}\, ,
\ee
where
\be
    \chi = \tau\, ,\quad z = \textrm{tan} \frac{\theta}{2}\, e^{i (\varphi - i \Omega \tau)}\, ,\quad \bar{z} = \textrm{tan} \frac{\theta}{2}\, e^{- i (\varphi - i \Omega \tau)}\, ,\\
\ee
with
\be
  f = 1\, ,\quad a = 0\, ,\quad c = L (1 + \textrm{cos}\, \theta) \, .\label{eq:def-c}
\ee

We can also define a new coordinate for later convenience:
\be
  \widetilde{\varphi} \equiv \varphi - i \Omega \tau\, .
\ee
Then,  the metric \eqref{eq:metric} becomes
\be\label{eq:metric-3}
  ds^2 = d\tau^2 + L^2 \Big[d\theta^2 + \textrm{sin}^2 \theta \, d\widetilde{\varphi}^2\Big]\, ,
\ee
which locally describes the direct product space $S^1 \times S^2$.  Hence,  the metric \eqref{eq:metric} of $S^1 \times$ (rotating $S^2$) can be obtained locally from the metric \eqref{eq:metric-3} of $S^1 \times$ (non-rotating $S^2$), if we replace the preriod of $\tau$ in \eqref{eq:OldPeriod},  i.e.  $\beta$,  with a new period
\be\label{eq:NewPeriod}
  \varphi \sim \varphi + 2 \pi\, ,\quad \tau \sim \tau - \frac{2 \pi i}{\Omega}\, ,
\ee
to preserve the regularity for $\varphi \in [0, \, 2\pi]$.  To manifest these symmetries in the coordinates $z$ and $\bar{z}$,  the $S^1$ direction $\tau$ in the metric \eqref{eq:metric} should have double periodicity
\be\label{eq:DoublePeriods}
  \tau \sim \tau + \mathfrak{p}\cdot \beta - \mathfrak{q}\cdot\frac{2 \pi i}{\Omega}\, ,\quad \varphi \sim \varphi + 2 \pi \mathfrak{q}\, ,
\ee
where $\beta$ and $2 \pi i / \Omega$ are two independent periods,  and $(\mathfrak{p},\, \mathfrak{q}) \in \mathbb{Z}^2$. Notice that $\Omega$ in \eqref{eq:DoublePeriods} can be either real or purely imaginary.

Following the framework of rigid supersymmetry, in order to define a pair of supercharges on the curved background spacetime \eqref{eq:metric}, we study solutions of the Killing spinor equations following from the gravitino variation in the new minimal off-shell supergravity:
\begin{align}
  (\nabla_\mu - i A_\mu) \zeta & = - \frac{1}{2} H \gamma_\mu \zeta - i V_\mu \zeta - \frac{1}{2} \epsilon_{\mu\nu\rho} V^\nu \gamma^\rho \zeta\, ,\label{eq:KillingSpEq-1}\\
  (\nabla_\mu + i A_\mu) \widetilde{\zeta} & = - \frac{1}{2} H \gamma_\mu \widetilde{\zeta} + i V_\mu \widetilde{\zeta} + \frac{1}{2} \epsilon_{\mu\nu\rho} V^\nu \gamma^\rho \widetilde{\zeta}\, ,\label{eq:KillingSpEq-2}
\end{align}
To completely determine the background, we assume the metric \eqref{eq:metric} and seek background fields $H$,  $A_\mu$, and $V_\mu$, verifying the above equations. This process has been sketched in various works \cite{Closset:2012ru,  Nian:2013qwa,  Nian:2019pxj},  and we will follow \cite{Nian:2019pxj} in this paper. The background fields $(H, \, A_\mu,\, V_\mu)$  satisfying \eqref{eq:KillingSpEq-1} and \eqref{eq:KillingSpEq-2} together with the curved-space metric \eqref{eq:metric} are a natural generalization of the original A-twisted background used in supersymmetric localization of 2d theories.

 In the frame given by
\be\label{eq:frame}
  e_1 = d\tau\, ,\quad e_z \equiv e_2 + i e_3 = c\, dz\, ,\quad e_{\bar{z}} \equiv e_2 - i e_3 = c\, d\bar{z},
\ee
the Killing spinors are
\be\label{eq:KillingSp-1}
  \zeta_\alpha = \sqrt{s} \left( \begin{array}{c}
  1 \\ 0
  \end{array} \right)\, ,\quad
  \zeta^\alpha = \sqrt{s} \left( \begin{array}{c}
  0 \\ -1
  \end{array} \right)\, .
\ee
\be\label{eq:KillingSp-2}
  \widetilde{\zeta}_\alpha = \frac{1}{\sqrt{s}} \left( \begin{array}{c}
  0 \\ 1
  \end{array} \right)\, ,\quad
  \widetilde{\zeta}^\alpha = \frac{1}{\sqrt{s}} \left( \begin{array}{c}
  1 \\ 0
  \end{array} \right)\, ,
\ee
where we make a special choice $s = e^{- 2 \tau (4 \Phi - \Omega)}$ with purely imaginary constants $\Phi$ and $\Omega$.  Correspondingly,  the background fields have the following expressions:
\begin{align}
\begin{split}
  H & = i \kappa\, ,\\
  A_1 = i (4 \Phi - \Omega) + \frac{3}{2} \kappa\, ,\quad A_2 & = \frac{i}{4 L} (z - \bar{z})\, ,\quad A_3 = \frac{1}{4 L} (z + \bar{z})\, ,\\
    V_1 = \kappa\, ,\quad V_2 & = 0\, ,\quad V_3 = 0\, ,
\end{split}
\end{align}
which always satisfy
\be
  A_1 - \frac{1}{2} V_1 + i H = i (4 \Phi - \Omega)\, .
\ee
We can particularize to\footnote{Note that our choice for the background fields, other than the metric,  differs from the one in Sec.~9.1 of \cite{Closset:2019hyt},  which they referred to as the SCI background; we call our choice the SCI-inspired background.  We plan to systematically explore the implications of such choices, following the framework put forward in \cite{Nian:2013qwa},  elsewhere.}:
\be\label{eq:FixGauge}
  A_1 - \frac{1}{2} V_1 = 0\quad\Rightarrow\quad H = 4 \Phi - \Omega\, .
\ee
For the Killing spinors to satisfy the anti-periodic boundary condition along $S^1$ requires
\be\label{eq:Constraint}
  \beta (4 \Phi - \Omega) = \pi i \quad (\text{mod } 2 \pi i)\, .
\ee
In this paper we choose $\beta (4 \Phi - \Omega) = \pi i$ for simplicity,  which together with \eqref{eq:FixGauge} fixes
\be\label{eq:Fixkappa}
  H = i \kappa = \frac{\pi i}{\beta}\, .
\ee

\section{The 2d Perspective of 3d SCFTs}\label{sec:2D}

The 3d background we discussed in detail in Sec.~\ref{sec:Bgd} is essentially the 3d $\mathcal{N}=2$ supersymmetric background discussed in \cite{Closset:2012ru,  Closset:2017zgf,Closset:2019hyt}.  It was shown in \cite{Closset:2017zgf} that the 3d $\mathcal{N}=2$ gauge theory can be supersymmetrically defined on a manifold, $\mathcal{M}_{g,  p}$,  the oriented circle bundle of degree $p$ over a closed Riemann surface of genus $g$,  $\Sigma_g$.  The advantage of this approach is that it acquires a topological nature.  Namely,  by compactifying the 3d theory on the circle $S_\beta^1$,  we obtain a 2d $\mathcal{N} = (2,2)$ theory,  whose low-energy effective theory is given by a Landau-Ginzburg theory with a twisted superpotential $\mathcal{W}$.  This is precisely the 2d effective theory that we are looking for.  It is  not really  a 2d theory, rather a 2d theory with infinitely many degrees of freedom, the theory remembers its 3d origin via these infinitely many degrees of freedom.

In this paper,  we will exploit a version of the 3d to 2d reduction that includes the sum over all KK modes,  which manifests itself in the expression for the twisted superpotential containing di-logarithmic functions $\textrm{Li}_2 (x)$,  as explained in, for example, \cite{Aharony:2017adm}.  The fact that we re-sum all the KK mode before taking the $\beta\to 0$ limit,  has important implications {\it vis-\`a-vis} direct 3d Cardy-like expansions considered in \cite{Choi:2019dfu,  GonzalezLezcano:2022hcf,  Bobev:2022wem, Bobev:2024mqw}.

The 2d $\mathcal{N} = (2,2)$ effective theory with the chiral multiplets $\Phi_i$ and the vector multiplet $\mathcal{V}$ is given by the action
\be
  S_{TFT} = \int_{\Sigma_g} d^2 x\, \sqrt{g} \left(- 2 f_{1 \bar{1} a} \frac{\partial \mathcal{W}}{\partial \sigma_a} + \widetilde{\Lambda}^a_{\bar{1}}\, \Lambda^b_1\, \frac{\partial^2 \mathcal{W}}{\partial \sigma_a\, \partial \sigma_b} \right) + \frac{i}{2} \int_{\Sigma_g} d^2 x\, \sqrt{g}\, \widetilde{\Omega} (\sigma)\, R\, ,
\ee
where $\mathcal{W}$ and $\widetilde{\Omega}$ are the twisted superpotential and the effective dilaton, respectively.  Here,  $\Lambda^b_1$,  $\widetilde{\Lambda}^a_{\bar{1}}$ and $f_{1 \bar{1} a}$ are components of a twisted chiral multiplet $\Sigma = -i D_- \widetilde{D}_+ \mathcal{V}$:
\be
  \Sigma = \left(\sigma,\, \Lambda_1,\, - \widetilde{\Lambda}_{\bar{1}},\,  -4\, f_{1 \bar{1}} \right)\, .
\ee

Given the 2d data composed of a twisted superpotential $\mathcal{W}$ and an effective dilaton $\widetilde{\Omega}$,  we can compute the partition function of the 3d $\mathcal{N}=2$ gauge theory on a curved background.  Indeed, the ingredients needed are: the handle-gluing operator,  the fibering operator,  and the flux operators (including the gauge magnetic flux and the flavor flux):
\begin{align}
  \mathcal{H} (u) & \equiv \textrm{det}_{ab} \left(\frac{\partial^2 \mathcal{W} (u)}{\partial u_a\, \partial u_b} \right)\cdot e^{2 \pi i\, \widetilde{\Omega} (u)}\, ,\\
  \mathcal{F} (u) & \equiv \textrm{exp} \left(2 \pi i \left(\mathcal{W} (u) - u_a\, \frac{\partial \mathcal{W} (u)}{\partial u_a} \right)\right)\, ,\\
  \Pi_a (u,  \nu) & \equiv \textrm{exp} \left(2 \pi i \frac{\partial \mathcal{W} (u,  \nu)}{\partial u_a} \right)\, ,\\
  \Pi_i (u,  \nu) & \equiv \textrm{exp} \left(2 \pi i \frac{\partial \mathcal{W} (u,  \nu)}{\partial \nu_i} \right)\, .
\end{align}
The topological nature of this kind of 3d partition function now manifests itself in the independence of the precise insertion points of the above operators,  which implies that the partition function can be expressed as
\be\label{eq:simplified Z}
  Z_{\mathcal{M}_{g,  p}} (\nu) = \sum_{\hat{u} \in \mathcal{S}_{BE}} \mathcal{F} (\hat{u},  \nu)^p\, \mathcal{H} (\hat{u},  \nu)^{g-1}\, \prod_i \Pi_i (\hat{u},  \nu)^{\mathfrak{n}_i}\, .
\ee
where $\mathfrak{n}_i$ denote the flavor magnetic fluxes,  and
\be\label{eq:def chemical potentials}
  u_a \equiv i \widetilde{R}_\beta (\sigma_a + i\, a_{0 a})\, ,\quad \nu_i \equiv i \widetilde{R}_\beta (m_i + i\, a^F_{0 i})\, .
\ee
Here,  $a_{0 a}$ and $a^F_{0 i}$ denote the holonomies of $G$ and $G_F$ along $S^1$ respectively,  and $\widetilde{R}_\beta$ is the radius of the $S^1$ circle.  In \eqref{eq:simplified Z},  $\mathcal{S}_{BE}$ stands for the set of the Bethe roots,  {\it i.e.},  the solutions to the Bethe Ansatz equations (BAE) expressed in terms of the color magnetic flux operators:
\be\label{eq:general BAE}
  \Pi_a (u,  \nu) = \textrm{exp} \left(2 \pi i \frac{\partial \mathcal{W} (u,  \nu)}{\partial u_a} \right) = 1\, .
\ee
The fact that the color magnetic flux operators $\Pi_a$ do not explicitly show up in the partition function \eqref{eq:simplified Z} can be explained as follows.  Schematically,  the partition function $Z_{\mathcal{M}_{g,  p}}$ can be expressed as a contour integral \cite{Closset:2017zgf}:
\be
  Z_{\mathcal{M}_{g,  p}} = \sum_{\mathfrak{m} \in \mathbb{Z}} \int_{\mathcal{C}} du\, \mathcal{J} (u)\, \Pi (u)^{\mathfrak{m}}\, ,
\ee
which contains a sum over the color magnetic fluxes.  By choosing a special Bethe Ansatz equation contour $\mathcal{C}_{BE}$,  the sum over the color magnetic fluxes can be evaluated as a geometrical series,  {\it i.e.},
\be
  Z_{\mathcal{M}_{g,  p}} = \int_{\mathcal{C}_{BE}} du\, \mathcal{J} (u)\, \frac{1}{1 - \Pi (u)}\, ,
\ee
which becomes a sum of the residues evaluated at the poles given by the Bethe Ansatz equation $\Pi (u) = 1$.  This procedure justifies the forms of the partition function \eqref{eq:simplified Z} and the Bethe Ansatz equations \eqref{eq:general BAE}.

\subsection{The ABJM Theory on $S^1 \times \textrm{(rotating  $S^2$)}$}

\begin{figure}[!htb]
      \begin{center}
        \includegraphics[width=0.45\textwidth]{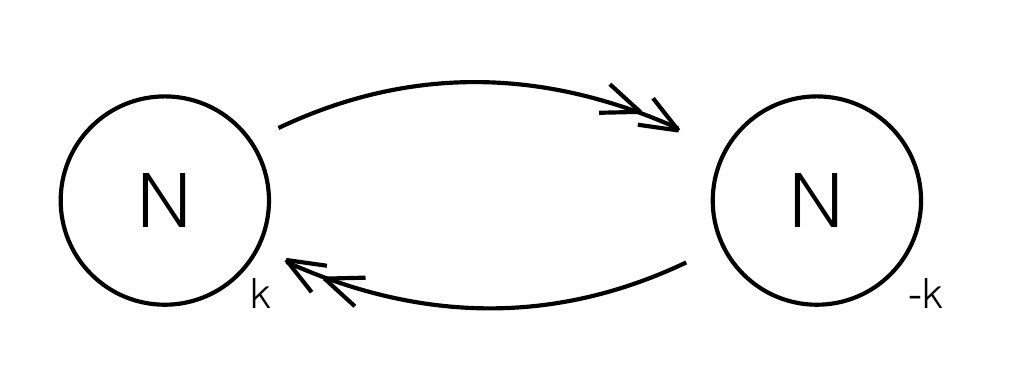}
        \caption{The quiver diagram of the 3d ABJM theory}
        \label{fig:ABJM}
      \end{center}
    \end{figure}

Let us consider the ABJM theory as a special example.  The ABJM theory is a 3d gauge theory with gauge group $U(N)_k \times U(N)_{-k}$ where the  Chern-Simons levels are $k$ and $-k$ and its  R-symmetry group is $SO(6) \cong SU(4)$ \cite{Aharony:2008ug}.  In 3d $\mathcal{N}=2$ language,  the ABJM theory has 2 vector multiplets of R-charge $2$,  whose gauge fields are $A_\mu$ and $\widetilde{A}_\mu$ for $U(N) \times U(N)$.  In addition,  the matter fields are
\be
  C_I = (A_1,\, A_2,\, \bar{B}^{\widetilde{1}},\, \bar{B}^{\widetilde{2}})\, ,\quad \Psi^I = (-\psi_2,\, \psi_1,\, - \bar{\chi}^{\widetilde{2}},\, \bar{\chi}^{\widetilde{1}})\, ,
\ee
which can be expressed as 4 chiral multiplets of R-charge $\frac{1}{2}$ as
\be
  (A_i, \psi_{i \alpha}) \in (N, \overline{N})\, ,\quad (B_{\widetilde{i}}, \chi_{\widetilde{i} \alpha}) \in (\overline{N}, N)\, ,
\ee
where $i = 1, 2$ and $\widetilde{i} = \widetilde{1}, \widetilde{2}$ are the indices of $SU(2) \times SU(2) \subset SU(4)_R$.  Schematically,  the ABJM theory can be formulated as a quiver gauge theory as shown in Fig.~\ref{fig:ABJM}.

Let us recall the charges of the field content of ABJM under its R-symmetry group,  $SO(6)_R$.  The three Cartans of $SO(6)_R$ are labelled by $h_1$, $h_2$ and $h_3$.  For the subgroup $SU(2) \times SU(2)$,  the two Cartans are chosen to be $\frac{1}{2} (h_1 \pm h_2)$.  The fields and their charges $h_{1, 2, 3}$ are listed in the Table \ref{R-charges} \cite{Bhattacharya:2008bja, Kim:2009wb}. 
\begin{table}[h!]
\centering
\begin{tabular}{c|ccc}
\hline
fields & $h_1$ & $h_2$ & $h_3$\\
\hline\\
$(A_1,\, A_2)$ & $\left(\frac{1}{2},\, - \frac{1}{2}\right)$ & $\left(\frac{1}{2},\, - \frac{1}{2}\right)$ & $\left(- \frac{1}{2},\, - \frac{1}{2}\right)$ \\
$(B_{\widetilde{1}},\, B_{\widetilde{2}})$ & $\left(\frac{1}{2},\, - \frac{1}{2}\right)$ & $\left(- \frac{1}{2},\, \frac{1}{2}\right)$ & $\left(- \frac{1}{2},\, - \frac{1}{2}\right)$ \\
$(\psi_{1 \pm},\, \psi_{2 \pm})$ & $\left(\frac{1}{2},\, - \frac{1}{2}\right)$ & $\left(\frac{1}{2},\, - \frac{1}{2}\right)$ & $\left(\frac{1}{2},\, \frac{1}{2}\right)$ \\
$(\chi_{\widetilde{1} \pm},\, \chi_{\widetilde{2} \pm})$ & $\left(\frac{1}{2},\, - \frac{1}{2}\right)$ & $\left(- \frac{1}{2},\, \frac{1}{2}\right)$ & $\left(\frac{1}{2},\, \frac{1}{2}\right)$ \\
\hline\\
$A_\mu,\, \widetilde{A}_\mu$ & $0$ & $0$ & $0$ \\
$\lambda_\pm,\, \widetilde{\lambda}_\pm$ & $0$ & $0$ & $-1$ \\
$\sigma,\, \widetilde{\sigma}$ & $0$ & $0$ & $0$ \\
\hline\\
$Q_\pm$ & $0$ & $0$ & $1$ \\
$S^\pm$ & $0$ & $0$ & $-1$
\end{tabular}
\caption{Charges of the fields in the ABJM theory\label{R-charges}}
\end{table}

In addition,  the representations for the 4 chiral multiplets of the ABJM theory are represented in Table \ref{ABJM-Rep}.
\begin{table}[h!]
\centering
\begin{tabular}{c|cc|ccccc}
\hline
fields & $U(N)_1$ & $U(N)_2$ & $SU(N_\chi)_{N_\chi=2}^{(1)}$ & $SU(N_\chi)_{N_\chi=2}^{(2)}$ & $U(1)_A$ & $U(1)_T$ & $U(1)_R$\\
\hline
{} & {} & {} & {} & {} & {} & {} & {} \\
$Q_1 = (A_1,\, \psi_1)$ & $N$ & $\overline{N}$ & $N_\chi$ & $1$ & $1$ & $0$ & $r/2$\\
$Q_2 = (A_2,\, \psi_2)$ & $N$ & $\overline{N}$ & $N_\chi$ & $1$ & $1$ & $0$ & $r/2$\\
$\widetilde{Q}_{\widetilde{1}} = (B_{\widetilde{1}},\, \chi_{\widetilde{1}})$ & $\overline{N}$ & $N$ & $1$ & $N_\chi$ & $1$ & $0$ & $r/2$\\
$\widetilde{Q}_{\widetilde{2}} = (B_{\widetilde{2}},\, \chi_{\widetilde{2}})$ & $\overline{N}$ & $N$ & $1$ & $N_\chi$ & $1$ & $0$ & $r/2$
\end{tabular}
\caption{The representations of the $\mathcal{N}=2$ chiral multiplets in ABJM theory \label{ABJM-Rep}}
\end{table}

Using the method of \cite{Closset:2017zgf},  we can collect the contributions from each chiral multiplet and vector multiplet including Chern-Simons contact terms.  The results of the twisted superpotential and the effective dilaton of the ABJM theory on $S^1 \times S^2$ are given by
\begin{align}
  (2 \pi i)^2\, \mathcal{W}^{ABJM} & = \sum_{a=1}^N \sum_{b=1}^N \Bigg[ \sum_{i=1}^{N_\chi = 2} \textrm{Li}_2 \left(x_a\, \widetilde{x}_b^{-1}\, y_i \right) + \sum_{j=1}^{N_\chi = 2} \textrm{Li}_2 \left(x_a^{-1}\, \widetilde{x}_b\, \widetilde{y}_j \right) \nonumber\\
  {} & \qquad\qquad + \sum_{i=1}^{N_\chi = 2} \frac{1}{2}\, \textrm{log}\, y_i \left(\textrm{log}\, y_i + 2 \pi i \right) + \sum_{j=1}^{N_\chi = 2} \frac{1}{2}\, \textrm{log}\, \widetilde{y}_j \left(\textrm{log}\, \widetilde{y}_j + 2 \pi i \right) \nonumber\\
  {} & \qquad\qquad + \frac{N_\chi}{2}\, \textrm{log}\, x_a (\textrm{log}\, x_a + 2 \pi i) + \textrm{log}\, q\, \textrm{log}\, x_a + \frac{N_\chi}{2}\, \textrm{log}\, \widetilde{x}_b (\textrm{log}\, \widetilde{x}_b + 2 \pi i) + \textrm{log}\, \widetilde{q}\, \textrm{log}\, \widetilde{x}_b \Bigg] \nonumber\\
  {} & \quad + \sum_{a=1}^N \frac{k}{2}\, \textrm{log}\, x_a \left(\textrm{log}\, x_a + 2 \pi i \right) - \sum_{b=1}^N \frac{k}{2}\, \textrm{log}\, \widetilde{x}_b \left(\textrm{log}\, \widetilde{x}_b + 2 \pi i \right) \, ,\label{eq:ABJM W}
\end{align}
\begin{align}
  \widetilde{\Omega}^{\text{ABJM}} & = \sum_{a=1}^N \sum_{b=1}^N \Bigg[ - \sum_{i=1}^{N_\chi = 2} \left(\frac{r_i / 2 - 1}{2 \pi i} \right) \textrm{log} (1 - x_a\, \widetilde{x}_b^{-1}\, y_i) - \sum_{j=1}^{N_\chi = 2} \left(\frac{\widetilde{r}_j / 2 - 1}{2 \pi i} \right) \textrm{log} (1 - x_a^{-1}\, \widetilde{x}_b\, \widetilde{y}_j) \nonumber\\
  {} & \qquad\qquad\quad + \frac{1}{N_\chi} \sum_{i=1}^{N_\chi = 2} \left(\frac{r_i / 2 - 1}{2 \pi i} \right) \textrm{log}\, y_i + \frac{1}{N_\chi} \sum_{j=1}^{N_\chi = 2} \left(\frac{\widetilde{r}_j / 2 - 1}{2 \pi i} \right) \textrm{log}\, \widetilde{y}_j \Bigg] \nonumber\\
  {} & \quad - \frac{1}{2 \pi i} \sum_{\substack{a,  b =1\\ a\neq b}}^N \textrm{log} (1 - x_a\, x_b^{-1}) - \frac{1}{2 \pi i} \sum_{\substack{a,  b=1\\ a\neq b}}^N \textrm{log} (1 - \widetilde{x}_a^{-1}\, \widetilde{x}_b)\, ,\label{eq:ABJM Omega}
\end{align}
where we have introduced the fugacities of the gauge and the flavor symmetries
\be
  x_a \equiv e^{2 \pi i u_a}\, , \quad \widetilde{x}_b \equiv e^{2 \pi i \widetilde{u}_b}\, , \quad y_i \equiv e^{2 \pi i \nu_i}\, , \quad \widetilde{y}_j \equiv e^{2 \pi i \widetilde{\nu}_j}\, ,
\ee
satisfying the constraints due to the $SU(N_\chi)^{(1)} \times SU(N_\chi)^{(2)}$ flavor symmetry
\be
  \prod_{i=1}^{N_\chi = 2} y_i = 1\, , \qquad \prod_{j=1}^{N_\chi = 2} \widetilde{y}_j = 1\, , \qquad \prod_{a=1}^N x_a = \prod_{b=1}^N \widetilde{x}_b\, .
\ee
These constraints correspond to a subset of the more general parameter space,  a choice that simplifies the computations significantly.  However,  the solutions hold for general fugacities at the leading order.  In the effective dilaton \eqref{eq:ABJM Omega},  we have introduced $r_i$ and $\widetilde{r}_j$ as the trial R-charges instead of the physical R-charge $r = 1$.

The Bethe roots are solutions to the Bethe Ansatz equations
\be
  \frac{\partial \mathcal{W}}{\partial u_a} = 0\, ,\quad \frac{\partial \mathcal{W}}{\partial \widetilde{u}_b} = 0\, , \qquad \forall a \in \{1, \cdots,  N \}\, ,\,\, \forall b \in \{1, \cdots,  N \}\, .
\ee
For the ABJM theory with the twisted superpotential given by \eqref{eq:ABJM W},  the Bethe Ansatz equations can be expressed as
\begin{align}
\begin{split}\label{eq:BAE for ABJM}
  (-1)^{N + k}\, x_a^k\, \frac{\prod_{b=1}^N \left(1 - \frac{\widetilde{x}_b}{x_a}\, \widetilde{y}_1 \right)\, \left(1 - \frac{\widetilde{x}_b}{x_a}\, \widetilde{y}_2 \right)}{\prod_{b=1}^N \left(1 - \frac{\widetilde{x}_b}{x_a}\, y_1^{-1} \right)\, \left(1 - \frac{\widetilde{x}_b}{x_a}\, y_2^{-1} \right)} & = 1\, ,\qquad  a \in \{1, \cdots,  N \}\, ,\\
  (-1)^{N + k}\, \widetilde{x}_b^k\, \frac{\prod_{b=1}^N \left(1 - \frac{\widetilde{x}_b}{x_a}\, \widetilde{y}_1 \right)\, \left(1 - \frac{\widetilde{x}_b}{x_a}\, \widetilde{y}_2 \right)}{\prod_{b=1}^N \left(1 - \frac{\widetilde{x}_b}{x_a}\, y_1^{-1} \right)\, \left(1 - \frac{\widetilde{x}_b}{x_a}\, y_2^{-1} \right)} & = 1\, ,\qquad  b \in \{1, \cdots,  N \}\, .
\end{split}
\end{align}
To simplify the computations,  we have chosen the fugacities $q$ and $\widetilde{q}$ such that
\be
  q^{N}\, \prod_{b=1}^N \widetilde{x}_b = 1\, ,\quad \widetilde{q}^N\, \prod_{a=1}^N x_a = 1\, .
\ee
We see that the Bethe Ansatz equations \eqref{eq:BAE for ABJM} for the 3d supersymmetric partition function only differ by a phase compared to the ones for the topologically twisted index \cite{Benini:2015eyy}.  Similar to \cite{Benini:2015eyy},  we can define a Bethe potential,  whose variation gives the Bethe Ansatz equations.  It was first observed in \cite{Hosseini:2016tor} that there exists a map between the large-$N$ Bethe potential and the $S^3$ free energy.  A similar map was found in \cite{Nian:2019pxj} between the large-$N$ Bethe potential and the $S^1 \times (\text{rotating } S^2)$ free energy.  The interplay between the large-$N$ limit and the Cardy-like expansion of the SCI was recently discussed in \cite{GonzalezLezcano:2022hcf} with the result that one needs first to take the large-$N$ limit to guarantee convergence of the Cardy expansion \cite{Closset:2019hyt}; here,  we have re-summed all the KK modes before taking any limits,  and that seems to have an important impact in the final results.

For the $\mathcal{M}_{g, p} = S^1 \times S^2$,  the partition function \eqref{eq:simplified Z} has contributions only from the handle-gluing operator and the flavor fluxes:
\be
   Z_{S^1 \times S^2}^{\text{ABJM}} = \sum_{\hat{u} \in \mathcal{S}_{BE}} \mathcal{H} (\hat{u},  \nu)^{-1}\, \prod_{i=1}^{N_\chi = 2} \Pi_i (\hat{u},  \nu_i)^{\mathfrak{n}_i}\, \prod_{j=1}^{N_\chi = 2} \widetilde{\Pi}_j (\hat{u},  \widetilde{\nu}_j)^{\widetilde{\mathfrak{n}}_j}\, .
\ee
Since we are interested in electrically charged rotating AdS$_4$ black holes without magnetic fluxes,  we set
\be
  \mathfrak{n}_i = \widetilde{\mathfrak{n}}_j = 0\, ,\quad \forall i,  j \in \{1,  2\}\, .
\ee
Hence,  for this case the partition function of the ABJM theory can be further simplified:
\begin{align}
  Z_{S^1 \times S^2}^{\text{ABJM}} & = \sum_{\hat{u} \in \mathcal{S}_{BE}} \mathcal{H} (\hat{u},  \nu)^{-1} \nonumber\\
  {} & = \sum_{\hat{u} \in \mathcal{S}_{BE}} \Bigg[\textrm{det}_{ab} \left(\frac{\partial^2 \mathcal{W} (u)}{\partial u_a\, \partial u_b} \right)\Bigg|_{u = \hat{u}} \cdot e^{2 \pi i\, \widetilde{\Omega} (\hat{u})} \Bigg]^{-1} \nonumber\\
  {} & = (2 \pi)^{2 N}\, \left(\frac{y_1^{\frac{r_1}{2} - 1}\, y_2^{\frac{r_2}{2} - 1} }{\widetilde{y}_1^{\frac{\widetilde{r}_1}{2} - 1}\, \widetilde{y}_2^{\frac{\widetilde{r}_2}{2} - 1} } \right)^{\frac{N^2}{2}}\cdot \sum_{\hat{u} \in \mathcal{S}_{BE}} \Bigg\{ \frac{1}{\textrm{det}\, \mathbb{B}}\, \prod_{b > a} \left(1 - \frac{x_b}{x_a} \right)^2\, \left(1 - \frac{\widetilde{x}_b}{\widetilde{x}_a} \right)^2 \nonumber\\
  {} & \qquad\qquad\qquad \cdot \prod_{a=1}^N \Bigg[ \frac{\widetilde{x}_a}{x_a} \prod_{i=1}^{N_\chi = 2} \left(1 - \frac{\widetilde{x}_a}{x_a}\, y_i^{-1} \right)^{\frac{r_i}{2} - 1}\cdot \prod_{j=1}^{N_\chi = 2} \left(1 - \frac{\widetilde{x}_a}{x_a}\, \widetilde{y}_j \right)^{\frac{\widetilde{r}_j}{2} - 1} \Bigg] \nonumber\\
  {} & \qquad\qquad\qquad \cdot \prod_{b>a} \Bigg[\prod_{i=1}^{N_\chi = 2} \left(1 - \frac{x_b}{\widetilde{x}_a}\, y_i \right)^{\frac{r_i}{2} - 1}\, \left(1 - \frac{\widetilde{x}_b}{x_a}\, y_i^{-1} \right)^{\frac{r_i}{2} - 1} \nonumber\\
  {} & \qquad\qquad\qquad\qquad\quad \cdot \prod_{j=1}^{N_\chi = 2} \left(1 - \frac{\widetilde{x}_b}{x_a}\, \widetilde{y}_j \right)^{\frac{\widetilde{r}_j}{2} - 1}\, \left(1 - \frac{x_b}{\widetilde{x}_a}\, \widetilde{y}_j^{-1} \right)^{\frac{\widetilde{r}_j}{2} - 1} \Bigg] \Bigg\}\, ,\label{eq:Z ABJM as TTI}
\end{align}
where the matrix $\mathbb{B}$ is defined as
\be
  \mathbb{B} \equiv \frac{\partial \left(B_1,\, \cdots,\, B_N,\, \widetilde{B}_1,\, \cdots,\, \widetilde{B}_N \right)}{\partial \left(u_1,\, \cdots,\, u_N,\, \widetilde{u}_1,\, \cdots,\, \widetilde{u}_N \right)}\, ,
\ee
with
\be
  B_a \equiv 2 \pi \frac{\partial \mathcal{W}}{\partial u_a}\, ,\quad \widetilde{B}_b \equiv - 2 \pi \frac{\partial \mathcal{W}}{\partial \widetilde{u}_b}\, ,\quad \forall a,  b \in \{1, \, \cdots,\,  N\}\, .
\ee
We see that the supersymmetric partition function of the ABJM theory via localization,  \eqref{eq:Z ABJM as TTI},  takes the same expression as the topologically twisted index \cite{Benini:2015eyy} up to an overall constant,  which is valid at finite $N$.  In the absence of magnetic fluxes,  the trial R-charges $r_i / 2$ and $\widetilde{r}_i / 2$ effectively play the role of the magnetic fluxes.  The coincidence of the Bethe Ansatz equations and the connection between the 3d supersymmetric partition function and the topologically twisted index possibly unveil a deep connection between these two physical problems.  Roughly speaking,  the existence of this relation is due to the identical way the topological twist and $Z_{S^1_\beta \times S^2}$ or the partition function on $S^1_\beta \times ({\rm rotating}\, S^2$) background fields enter the Killing spinor equation,  which will be discussed in full detail elsewhere \cite{Amariti:2023}.  Recently,  a similar relation between the 3d superconformal index (SCI) and the 3d topologically twisted index (TTI) has been observed in the Cardy-like expansion \cite{GonzalezLezcano:2022hcf,  Bobev:2022wem, Bobev:2024mqw}. A corresponding relation between the SCI and the TTI should also exist and result from rewriting the 3d SCI as a particular 3d supersymmetric partition function in the rigid supersymmetric backgrounds.

We can now proceed to the large-$N$ limit.  Following the same steps as in \cite{Benini:2015eyy},  we can evaluate the supersymmetric partition function of the ABJM theory on the special background described in Sec.~\ref{sec:Bgd} at the leading order in large $N$.  The result is
\be\label{eq:A-twist part fct ABJM}
  \textrm{log}\, Z_{S^1 \times S^2}^{\text{ABJM}} = - \frac{\sqrt{2}}{3}\, k^{\frac{1}{2}}\, N^{\frac{3}{2}}\, \sqrt{\Delta_1 \Delta_2 \Delta_3 \Delta_4} \sum_{I = 1}^4 \frac{R_I / 2}{\Delta_I}\, ,
\ee
where $\Delta_I$ are the chemical potentials defined by $y_I \equiv e^{i \Delta_I}$ with
\be
  y_I \equiv \left\{
  \begin{aligned}
    & y_{i = I}\, , & \textrm{for } I = 1,\, 2\, ; \\
    & \widetilde{y}_{j = I - 2}\, , & \textrm{for } I = 3,\, 4\, ,
  \end{aligned} \right. \qquad \text{and}\quad R_I \equiv \left\{
  \begin{aligned}
    & r_{i = I}\, , & \textrm{for } I = 1,\, 2\, ; \\
    & \widetilde{r}_{j = I - 2}\, , & \textrm{for } I = 3,\, 4\, .
  \end{aligned} \right.
\ee
The trial R-charges and the chemical potentials are related by $\Delta_I = \pi R_I/2$,  subject to the constraints
\be
  \sum_{I=1}^4 \Delta_I = 2 \pi\, .
\ee
We can first choose $R_I$ to minimize the partition function:
\be
  \left[\textrm{log}\, Z_{S^1 \times S^2}^{\text{ABJM}}\right]_{\text{min}} = - \frac{\sqrt{2}}{3 \pi}\, k^{\frac{1}{2}}\, N^{\frac{3}{2}}\, \sqrt{\Delta_1 \Delta_2 \Delta_3 \Delta_4}\, .
\ee
Next, let us recover the $\beta$-dependence by using the relations $\Delta_I\,\, \propto\,\, i \beta m_I$. Hence,  the equation above implies that $\left[\textrm{log}\, Z_{S^1 \times S^2}^{\text{ABJM}}\right]_{\text{min}} \propto\,\, \beta^2$.  We have seen in Eqs.~\eqref{eq:OldPeriod} and \eqref{eq:NewPeriod} that the metric of $S^1 \times$ (rotating $S^2$) can be obtained locally from the metric of $S^1 \times$ (non-rotating $S^2$) by changing the period of $\tau$ from $\beta$ to a new period $- 2 \pi i / \Omega$.  Therefore,  as a trick of introducing the angular velocity,  we use $(- 2 \pi i / \Omega)\, \beta$ to replace $\beta^2$.  This treatment turns out to give $\left[\textrm{log}\, Z_{S^1 \times S^2}^{\text{ABJM}}\right]_{\text{min}}$ the correct leading-order scaling in the Cardy-like limit. After this modification,  the result becomes
\begin{align}
  \left[\textrm{log}\, Z_{S^1 \times (\textrm{rot.} S^2)}^{\text{ABJM}}\right]_{\text{min}} & = - \frac{\sqrt{2}}{3 \pi}\, k^{\frac{1}{2}}\, N^{\frac{3}{2}}\, \left(- \frac{2 \pi i}{\beta \Omega} \right)\, \sqrt{\Delta_1 \Delta_2 \Delta_3 \Delta_4} \nonumber\\
  {} & = \frac{4 \sqrt{2}\, i}{3\, \widetilde{\omega}}\, k^{\frac{1}{2}}\, N^{\frac{3}{2}}\, \sqrt{\Delta_1 \Delta_2 \Delta_3 \Delta_4} \nonumber\\
  {} & = - \frac{4 \sqrt{2}\, i}{3\, \widetilde{\omega}}\, k^{\frac{1}{2}}\, N^{\frac{3}{2}}\, \sqrt{(i \Delta_1) (i \Delta_2) (i \Delta_3) (i \Delta_4)} \nonumber\\
  {} & = - \frac{4 \sqrt{2}\, i}{3\, \widetilde{\omega}}\, k^{\frac{1}{2}}\, N^{\frac{3}{2}}\, \sqrt{\widetilde{\Delta}_1\, \widetilde{\Delta}_2\, \widetilde{\Delta}_3\, \widetilde{\Delta}_4}\, ,\label{eq:AdS4 SBH from TTI}
\end{align}
where we have defined $\widetilde{\omega} \equiv 2 \beta \Omega$ and $\widetilde{\Delta}_I \equiv i \Delta_I$,  which in the Cardy-like limit $\widetilde{\omega} \sim \beta \to 0$ satisfy
\be
  \sum_{I=1}^4 \widetilde{\Delta}_I - \widetilde{\omega} = 2 \pi i + \mathcal{O} (\beta^2)\, .
\ee
This condition is consistent with purely imaginary-valued $\Omega$ and $\widetilde{\omega}$, and it originates from the one for the topologically twisted index,  $\sum_{I=1}^4 \Delta_I = 2 \pi$.  Since it was found in \cite{Benini:2015eyy} that there exist solutions to the Bethe Ansatz equations for $\sum_{I=1}^4 \Delta_I = 2 \pi$ and $\sum_{I=1}^4 \Delta_I = 6 \pi$,  but not for $\sum_{I=1}^4 \Delta_I = 4 \pi$.  We can conjecture a more general constraint for the chemical potentials:
\be\label{eq:constraint from TTI}
  \sum_{I=1}^4 \widetilde{\Delta}_I - \widetilde{\omega} = 2 \pi i\quad (\textrm{mod}\,\, 4 \pi i)\, .
\ee
This relation is exactly the BPS condition \eqref{eq:Constraint},  which guarantees the existence of Killing spinors.  Therefore,  we recover the same entropy function of the electrically charged BPS AdS$_4$ black hole, \eqref{eq:AdS4 SBH from TTI}, with the constraint \eqref{eq:constraint from TTI} from the 2d perspective and mapping the supersymmetric partition function on a particular rigid supersymmetric background given in Sec.~\ref{sec:Bgd} to the topologically twisted index.  For the AdS$_5$ BPS black holes,  a constraint similar to \eqref{eq:constraint from TTI} exists,  which has been used to discuss the supersymmetric index on the second sheet in \cite{Cassani:2021fyv}.


\subsection{The Mass Deformed ABJM Theory}

We now consider the mass-deformed ABJM (mABJM) theory which can be obtained from the ABJM theory by introducing a mass term deformation in the superpotential \cite{Warner:1983vz,  Warner:1983du,  Bobev:2018uxk}:
\be
  \Delta W  \sim \textrm{Tr} \left(T^{(1)}\, A_1 \right)^2\,.
\ee
The new addition to the superpotential fixes the R-charge of the chiral multiplet $A_1$ to be $r_1 / 2 = 1$,  because the R-charge of the monopole operator $T^{(1)}$ can be chosen to be vanishing (see \cite{Bobev:2018uxk} and the references therein).  Hence,  we should fix the trial R-charge for $A_1$,  {\it i.e.},
\be
  \frac{R_1}{2} = 1\, ,
\ee
while keeping the other 3 trial R-charges free subject to the sum of the R-charges:
\be
  \frac{R_2}{2} + \frac{R_3}{2} + \frac{R_4}{2} = 1\, .
\ee
Correspondingly,  the chemical potentials satisfy
\be
  \Delta_1 = \pi\, ,\quad \Delta_2 + \Delta_3 + \Delta_4 = \pi\, .
\ee

From the 2d effective theory point of view,  the mass deformed ABJM theory has the same twisted superpotential $\mathcal{W}$ and the effective dilaton $\widetilde{\Omega}$ as the original ABJM theory.  The only modification originates from the constraints on the trial R-charges and the chemical potentials discussed above.  As before,  if we identify the R-charges $R_I / 2$ in the 3d supersymmetric partition function defined on the rigid supersymmetric background for $S^1\times S^2$ and the magnetic fluxes $\mathfrak{n}_I$ in the topologically twisted index,  these two quantities take the same expression.  Hence,  we can repeat the same steps as in the previous subsection to compute the 3d supersymmetric partition function on the rigid supersymmetric background for the mass-deformed ABJM theory.  The result is as \eqref{eq:A-twist part fct ABJM} but with a special choice of the R-charges:
\be
  \textrm{log}\, Z_{S^1 \times S^2}^{\textrm{mABJM}} = - \frac{\sqrt{2}}{3}\, N^{\frac{3}{2}}\, \sqrt{\pi \Delta_2 \Delta_3 \Delta_4} \left( \frac{1}{\pi} + \frac{R_2 / 2}{\Delta_2} + \frac{R_3 / 2}{\Delta_3} + \frac{R_4 / 2}{\Delta_4} \right)\, ,
\ee
where we set the Chern-Simons level $k$ to be $1$.  For $S^1 \times S^2$ without rotation,  we do not need to modify the period of $S^1$ to incorporate the angular velocity.

Equivalently,  we can also study the topologically twisted index with the following magnetic fluxes
\be\label{eq:Constraint for mABJM}
  \mathfrak{n}_1 = 1\, ,\quad \mathfrak{n}_2 + \mathfrak{n}_3 + \mathfrak{n}_4 = 1\, .
\ee
This constraint is obtained in the following way.  The mass deformation breaks the global symmetry from $SO(8)$ to $SU(3) \times U(1)^W_R$.  The symmetry breaking is imposed by the condition
\be
  n^{(m)} \equiv n_1 - n_2 - n_3 - n_4 = 0\, ,
\ee
and the topological twist along the new R-symmetry leads to
\be
  n^{(R)} \equiv \frac{1}{2} (3 n_1 + n_2 + n_3 + n_4) = - 2 (\mathfrak{g} - 1)\, ,
\ee
with the genus $\mathfrak{g} = 0$ for the sphere $S^2$.  Solving these two conditions,  we obtain the constraint \eqref{eq:Constraint for mABJM}.  Notice that we choose an opposite sign convention compared to \cite{Bobev:2018uxk}.

At leading order in large $N$,  the topologically twisted index becomes
\begin{align}
  \textrm{log}\, Z_{S^1 \times S^2}^{\textrm{mABJM}} & = - \frac{\sqrt{2}}{3}\, N^{\frac{3}{2}}\, \sqrt{\pi \Delta_2 \Delta_3 \Delta_4} \left( \frac{1}{\pi} + \frac{\mathfrak{n}_2}{\Delta_2} + \frac{\mathfrak{n}_3}{\Delta_3} + \frac{\mathfrak{n}_4}{\Delta_4} \right) \nonumber\\
  {} & = - \frac{\sqrt{2} \pi}{3}\, N^{\frac{3}{2}}\, \sqrt{ \frac{\Delta_2}{\pi}\,  \frac{\Delta_3}{\pi}\, \frac{\Delta_4}{\pi}} \left( 1 + \frac{\mathfrak{n}_2}{\Delta_2 / \pi} + \frac{\mathfrak{n}_3}{\Delta_3 / \pi} + \frac{\mathfrak{n}_4}{\Delta_4 / \pi} \right)\, ,\label{eq:A-twist part fct mABJM}
\end{align}
subject to the constraint
\be
  \Delta_2 + \Delta_3 + \Delta_4 = \pi\, .
\ee
Note that we have adopted slightly different conventions compared to \cite{Bobev:2018uxk}.  Here,  $\mathfrak{n}_I$ are identified with $- \mathfrak{n}_I$ in \cite{Bobev:2018uxk},  while $\Delta_{2, 3, 4} / \pi$ can be identified with $\Delta_{2,3,4}$ in \cite{Bobev:2018uxk}.  With these identifications,  the result is the same as the topologically twisted index for the mABJM theory \cite{Bobev:2018uxk}.

\subsection{The AdS$_4$ Black Hole in Massive IIA Supergravity} 

\begin{figure}[!htb]
      \begin{center}
        \includegraphics[width=0.27\textwidth]{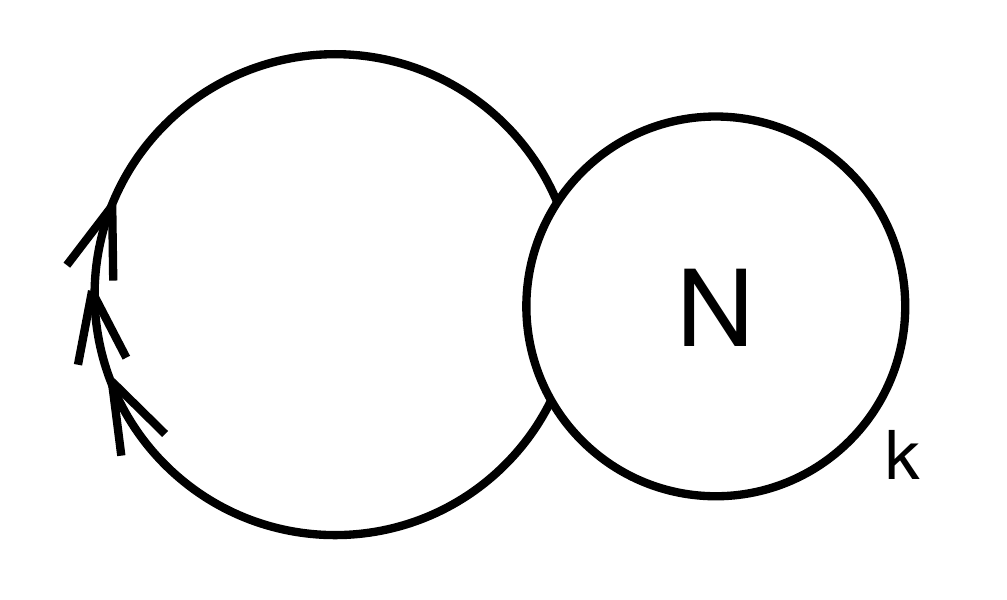}
        \caption{The quiver gauge theory dual to AdS$_4$ black hole in massive IIA supergravity}
        \label{fig:IIA}
      \end{center}
    \end{figure}

There is an interesting class of AdS$_4$ gravity solutions that can be lifted into massive IIA supergravity \cite{Romans:1985tz,  Guarino:2015jca},  and their free energy scales as $\sim N^{5/3}$ in the large-$N$ limit.  This class of theories can be obtained from $N$ D2-branes probing a Calabi-Yau threefold singularity in the presence of a non-zero Romans mass and has 3d dual field theory descriptions \cite{Guarino:2015jca,  Fluder:2015eoa}.  A microscopic foundation for the entropy of such AdS$_4$ black holes in massive IIA theories was provided in \cite{Hosseini:2017fjo,  Benini:2017oxt},  and the sub-leading logarithmic corrections to the Bekenstein-Hawking entropy were discussed in \cite{Liu:2018bac}.  The dual field theory content can be illustrated as a quiver gauge theory shown in Fig.~\ref{fig:IIA},  which consists of 1 vector multiplet and 3 chiral multiplets in the adjoint representation. More precisely, the representations of the chiral multiplets are listed in Table~\ref{tab:representations IIA}.

\begin{table}[h!]
\centering
\begin{tabular}{c|c|cccc}
\hline
fields & $U(N)$ & $SU(N_\chi)_{N_\chi = 3}$ & $U(1)_R$\\
\hline
{} & {} & {} & {} & {} & {} \\
$\phi_1$ & $(N,  \overline{N})$ & $N_\chi$ & $r/2$\\
$\phi_2$ & $(N,  \overline{N})$ & $N_\chi$ & $r/2$\\
$\phi_3$ & $(N,  \overline{N})$ & $N_\chi$ & $r/2$
\end{tabular}
\caption{The representations of the dual of the AdS$_4$ black hole in massive IIA theory}\label{tab:representations IIA}
\end{table}

From the 2d perspective,  the twisted superpotential is
\begin{align}
  (2 \pi i)^2\, \mathcal{W}\,^{\text{IIA}} & = \sum_{a=1}^N \sum_{b=1}^N \Bigg[ \sum_{I=1}^{N_\chi = 3} \textrm{Li}_2 \left( x_a^{-1} x_b y_I \right) + \sum_{I=1}^{N_\chi = 3} \frac{1}{2}\, \textrm{log}\, y_I \left(\textrm{log}\, y_I + 2 \pi i \right) \nonumber\\
  {} & \qquad\qquad + \frac{N_\chi}{2}\, \textrm{log}\, x_a \left(\textrm{log}\, x_a + 2 \pi i \right) + \textrm{log}\, q\, \textrm{log}\, x_a \Bigg] \nonumber\\
  {} & \quad + \sum_{a=1}^N \frac{k}{2}\, \textrm{log}\, x_a \left(\textrm{log}\, x_a + 2 \pi i \right)\, ,
\end{align}
and the effective dilaton is
\begin{align}
  \widetilde{\Omega}^{\text{IIA}} & = \sum_{a=1}^N \sum_{b=1}^N \Bigg[- \sum_{I=1}^{N_\chi = 3} \left(\frac{r_I / 2 - 1}{2 \pi i} \right)\, \textrm{log} \left(1 - x_a^{-1}\, x_b\, y_I \right) + \frac{1}{N_\chi} \sum_{I=1}^{N_\chi = 3} \left(\frac{r_I / 2 - 1}{2 \pi i} \right)\, \textrm{log}\, y_I \Bigg] \nonumber\\
  {} & \quad - \frac{1}{2 \pi i} \sum_{\substack{a,  b =1\\ a\neq b}}^N \textrm{log} (1 - x_a\, x_b^{-1})\, .
\end{align}
The corresponding Bethe Ansatz equations are
\be
  \frac{\partial \mathcal{W}^{\text{IIA}}}{\partial u_a} = 0\, ,
\ee
or more explicitly,
\be
  (-1)^{N + k}\, x_a^k \prod_{a \neq b=1}^N \prod_{I=1}^{N_\chi = 3} \frac{x_a - x_b\, y_I}{x_b - x_a\, y_I} = 1\, ,\qquad \forall a \in \{1, \,\cdots, \, N \}\, ,
\ee
where we have chosen the fugacity $q$ such that $q^N\, \prod_{b=1}^N x_b = 1$.  These Bethe Ansatz equations are up to a phase the same as the ones for the topologically twisted index \cite{Hosseini:2017fjo,  Benini:2017oxt}.

In the absence of magnetic fluxes,  {\it i.e.},  $\mathfrak{n}_I = 0,\,\, \forall I \in \{1,\, \cdots,  N_\chi\}$,  the 3d supersymmetric partition function $Z_{S^1 \times S^2}$ only receives contribution from the handle-gluing operator,  and the result is
\begin{align}
  Z_{S^1 \times S^2}^{\text{IIA}} & = \sum_{\hat{u} \in \mathcal{S}_{BE}} \mathcal{H} (\hat{u},  \nu)^{-1} \nonumber\\
  {} & = \sum_{\hat{u} \in \mathcal{S}_{BE}} \Bigg[\textrm{det}_{ab} \left(\frac{\partial^2 \mathcal{W} (u)}{\partial u_a\, \partial u_b} \right)\cdot e^{2 \pi i\, \widetilde{\Omega} (u)} \Bigg]^{-1} \nonumber\\
  {} & = (2 \pi)^N\cdot \prod_{I=1}^{N_\chi = 3} \frac{y_I^{N^2 (1 - r_I / 2)/3}}{(1 - y_I)^{N (1 - r_I / 2)}} \sum_{\hat{u} \in \mathcal{S}_{BE}} \frac{1}{\textrm{det}\, \mathbb{B}} \prod_{a \neq b} \left(1 - \frac{x_a}{x_b} \right)\cdot \prod_{I=1}^{N_\chi = 3} \left(1 - y_I\, \frac{x_a}{x_b} \right)^{\frac{r_I}{2} - 1}\, ,\label{eq:Z IIA as TTI}
\end{align}
where the matrix $\mathbb{B}$ is defined as
\be
  \mathbb{B} \equiv \frac{\partial \left(B_1,\, \cdots,\, B_N \right)}{\partial \left(u_1,\, \cdots,\, u_N \right)}\, ,\quad \text{with}\,\, B_a \equiv 2 \pi \frac{\partial \mathcal{W}}{\partial u_a}\, , \quad \forall a \in \{1, \,\cdots,\,  N\}\, .
\ee
If we make the identification $r_I / 2 = \mathfrak{n}_I$ or $r_I / 2 = - \mathfrak{n}_I$, \footnote{For the magnetic fluxes $\mathfrak{n}_I$ of flavor symmetry,  the conventions in \cite{Hosseini:2017fjo,  Benini:2017oxt} differ by a sign.} then up to an overall constant, the 3d supersymmetric partition function \eqref{eq:Z IIA as TTI} the background for $S^1\times ({\rm rotating})S^2$ is the same as the topologically twisted index for the field theory dual of the AdS$_4$ black hole in massive IIA supergravity \cite{Hosseini:2017fjo,  Benini:2017oxt}.  It is worth remarking that this relation between the supersymmetric partition function on the particular $S_\beta^1\times ({\rm rotating}) S^2$ background and the TTI is true at finite $N$.

Following the same procedure as in \cite{Hosseini:2017fjo,  Benini:2017oxt},  we can evaluate the 3d supersymmetric partition function in the large-$N$ limit,  and the result at the leading order is
\be\label{eq:A-twist part fct IIA}
  \textrm{log}\, Z_{S^1 \times S^2}^{\text{IIA}} = - \frac{3^{7/6}}{5 \times 2^{5/3}} \left(1 - \frac{i}{\sqrt{3}} \right)\, k^{1/3}\, N^{5/3}\, \left(\Delta_1\, \Delta_2\, \Delta_3 \right)^{2/3}\, \sum_{I=1}^3 \frac{\mathfrak{n}_I}{\Delta_I}\, ,
\ee
where $\mathfrak{n}_I = \frac{r_I}{2}$,  and $\Delta_I$ are the trial R-charges as chemical potentials defined by $y_I \equiv e^{i \Delta_I}$.  They satisfy
\be
  \sum_{I=1}^{N_\chi = 3} \mathfrak{n}_I \equiv \sum_{I=1}^{N_\chi = 3} \frac{r_I}{2} = 2\, ,\quad \sum_{I=1}^{N_\chi = 3} \Delta_I = 2 \pi\quad (\text{mod } 6 \pi)\, .
\ee

\subsection{The Family of Theories with the Scaling $\sim N^{5/3}$}

The field theory dual of the AdS$_4$ black holes in massive IIA theory has been generalized to a family of theories,  whose free energies in the large-$N$ limit scale as $\sim N^{5/3}$ \cite{Fluder:2015eoa}.  The large-$N$ indices of this class of theories have been computed in \cite{Hosseini:2022vho} with some new insights.

This class of theories has a gauge group $\mathcal{G} = \prod_{s=1}^{|\mathcal{G}|} U(N)_s$,  with a Chern-Simons level $k_s$ for $U(N)_s$,  and the matter content can be chiral multiplets in bi-fundamental and adjoint representations with $\sum k_s \neq 0$.  Ref.~\cite{Hosseini:2022vho} has computed the refined topologically twisted index with an equivariant parameter $\epsilon$ for this class of theories.  Here,  we set the refinement parameter $\epsilon = 0$ for simplicity.

Consider a generic theory in this class with $N_\chi$ chiral multiplets labelled by $I \in \{1,\, \cdots,\, N_\chi \}$ and $|\mathcal{G}|$ nodes of $U(N)$ labelled by $s \in \{1,\, \cdots,\, |\mathcal{G}|\}$.  From the 2d perspective,  the twisted superpotential of this theory is
\begin{align}
  (2 \pi i)^2\, \mathcal{W}^{\,\text{(5/3)}} & = \sum_{I=1}^{N_\chi} \sum_{a = 1}^N \sum_{b =1}^N \Bigg[ \textrm{Li}_2 \left( \left(x^{(s_I^+)}_{a} \right)^{-1} x^{(s_I^-)}_{b}\, y_I \right) + \frac{1}{2}\, \textrm{log}\, y_I \left(\textrm{log}\, y_I + 2 \pi i \right) \nonumber\\
  {} & \qquad\qquad\qquad\qquad + \frac{1}{2}\, \textrm{log}\, x^{(s_I^+)}_{a} \left(\textrm{log}\, x^{(s_I^+)}_{a} + 2 \pi i \right) + \frac{1}{N_\chi} \textrm{log}\, q\, \textrm{log}\, x^{(s_I^+)}_{a} \nonumber\\ 
  {} & \qquad\qquad\qquad\qquad + \frac{1}{2}\, \textrm{log}\, x^{(s_I^- \neq s_I^+)}_{b} \left(\textrm{log}\, x^{(s_I^- \neq s_I^+)}_{b} + 2 \pi i \right) + \frac{1}{N_\chi} \textrm{log}\, q\, \textrm{log}\, x^{(s_I^- \neq s_I^+)}_{b} \Bigg] \nonumber\\
  {} & \quad + \sum_{s=1}^{|\mathcal{G}|} \sum_{a =1}^N \frac{k_s}{2}\, \textrm{log}\, x^{(s)}_{a} \left(\textrm{log}\, x^{(s)}_{a} + 2 \pi i \right)\, ,
\end{align}
where $(s_I^+,\, s_I^-)$ with $I\in \{1,\, \cdots,\, N_\chi \}$ labels the $s_I^+$-th and the $s_I^-$-th nodes of $U(N)$ ($s_I^+,\,  s_I^- \in \{1,\, \cdots,\, |\mathcal{G}| \}$) connected by the $I$-th flavor,  and the corresponding fugacities of these two $U(N)$'s are denoted by $x^{(s_I^+)}$ and $x^{(s_I^-)}$ respectively.  For each $U(N)$ node,  its Cartan generators are labelled by $a$ (or $b$) with $a,\, b \in \{1,\, \cdots,\, N \}$.

The effective dilaton is
\begin{align}
  \widetilde{\Omega}^{\,\text{(5/3)}} & = \sum_{I=1}^{N_\chi} \sum_{a = 1}^N \sum_{b =1}^N \Bigg[- \left(\frac{r_I / 2 - 1}{2 \pi i} \right)\, \textrm{log} \left(1 - \left(x^{(s_I^+)}_{a}\right)^{-1}\, x^{(s_I^-)}_{b}\, y_I \right) + \frac{1}{N_\chi} \left(\frac{r_I / 2 - 1}{2 \pi i} \right)\, \textrm{log}\, y_I \Bigg] \nonumber\\
  {} & \quad - \frac{1}{2 \pi i} \sum_{s=1}^{|\mathcal{G}|} \sum_{\substack{a, \, b =1\\ a\neq b}}^N \textrm{log} \left(1 - x^{(s)}_{a}\, \left(x^{(s)}_{b}\right)^{-1} \right)\, .
\end{align}

The Bethe Ansatz equations are given by
\be
  \frac{\partial \mathcal{W}^{\,\text{(5/3)}}}{\partial u_a^{(s)}} = 0\, ,\quad \forall s \in \{1,\, \cdots,\, |\mathcal{G}| \}\, ,\,\, \forall a \in \{1,\, \cdots,\, N \}\, .
\ee
Here,  we skip the lengthy explicit expressions of the Bethe Ansatz equations.

In the absence of magnetic fluxes, the 3d supersymmetric partition function is
\begin{align}
  {}\, & Z_{S^1 \times S^2}^{\, \text{(5/3)}} \nonumber\\
  =\, & \sum_{\hat{u} \in \mathcal{S}_{BE}} \mathcal{H} (\hat{u},  \nu)^{-1} = \sum_{\hat{u} \in \mathcal{S}_{BE}} \Bigg[\textrm{det}_{ab} \left(\frac{\partial^2 \mathcal{W} (u)}{\partial u_a\, \partial u_b} \right)\cdot e^{2 \pi i\, \widetilde{\Omega} (u)} \Bigg]^{-1} \nonumber\\
  =\, & (2 \pi)^{|\mathcal{G}| N}\cdot \prod_{I=1}^{N_\chi} y_I^{\frac{N^2}{N_\chi} (1 - \frac{r_I}{2})} \sum_{\hat{u} \in \mathcal{S}_{BE}} \frac{1}{\textrm{det}\, \mathbb{B}} \left[\prod_{s=1}^{|\mathcal{G}|} \prod_{a \neq b} \left(1 - \frac{x^{(s)}_a}{x^{(s)}_b} \right)\right] \cdot \left[\prod_{I=1}^{N_\chi} \prod_{a,  b=1}^N \left(1 - y_I\, \frac{x^{(s_I^-)}_b}{x^{(s_I^+)}_a} \right)^{\frac{r_I}{2} - 1}\right]\, .\label{eq:Z 5/3 as TTI}
\end{align}
For $|\mathcal{G}| = 1$ and $N_\chi = 3$, the general partition function \eqref{eq:Z 5/3 as TTI} reduces to \eqref{eq:Z IIA as TTI} in the previous example.

Following the similar procedure of the large-$N$ topologically twisted index \cite{Hosseini:2022vho},  we can obtain the leading term of the large-$N$ 3d supersymmetric partition function:
\be\label{eq:A-twist logZ 5/3 large-N}
  \textrm{log}\, Z_{S^1 \times S^2}^{\text{(5/3)}} = - \frac{3^{7/6}}{10} \left(1 - \frac{i}{\sqrt{3}} \right)\, \left(\frac{k}{\sum_{I=1}^{N_\chi} g_+ (\Delta_I)} \right)^{\frac{1}{3}}\, N^{\frac{5}{3}}\, \left(\sum_{I=1}^{N_\chi} (\mathfrak{n}_I - 1)\, g'_+ (\Delta_I) + |\mathcal{G}|\, \frac{\pi^2}{3}\right)\, ,
\ee
where
\be
  k \equiv \sum_{s=1}^{|\mathcal{G}|} k_s\, ,\quad g_+ (x) \equiv \frac{x^3}{6} - \frac{\pi}{2}\, x^2 + \frac{\pi^2}{3}\, x\, ,
\ee
\be
  \mathfrak{n}_I = \frac{r_I}{2}\quad \text{with } \sum_{I=1}^{N_\chi} \mathfrak{n}_I = 2\, ,\quad \text{and}\quad \sum_{I=1}^{N_\chi} \Delta_I = 2 \pi \quad \text{with } y_I \equiv e^{i \Delta_I}\, .
\ee
In fact,  \eqref{eq:A-twist logZ 5/3 large-N} can also be written into a more inspiring form $\sim c_r / a^{1/3}$ with a 4d anomaly coefficient $a$ and a 2d central charge $c_r$ \cite{Hosseini:2016cyf},  which suggests the 4d origin of this family of 3d SCFTs.  In particular,  for the 4d ``parent'' SCFTs from D3-branes at toric singularities,  $\textrm{log}\, Z_{S^1 \times S^2}^{\text{(5/3)}}$ in \eqref{eq:A-twist logZ 5/3 large-N} is proportional to $\frac{C_{IJK}\, \mathfrak{n}_I \Delta_J \Delta_K}{(C_{IJK} \Delta_I \Delta_J \Delta_K)^{1/3}}$,  where $C_{IJK} = \textrm{Tr} (Q_I Q_J Q_K)$ is the anomaly coefficient which can also be obtained from the triangle connecting three vertices on the toric diagram \cite{Butti:2005vn} and was discussed holographically in \cite{Benvenuti:2006xg}.  For this special class of theories,  the constraints $\sum_I \Delta_I =2 \pi$  and  $\sum_I \mathfrak{n}_I =2$ become natural,  where $i$ runs over the whole set of $U(1)$ symmetries in the GLSM associated to the CY$_3$.

\section{The 3d SCFTs from 3d Perspective}\label{sec:3D}

In this section,  we consider 3d SCFTs from the 3d perspective.  We will see that the real and the imaginary parts of the free energy on $S^1_\beta \times S^2$ are related to the coefficients of the current-current 2-point correlators,  which provides an alternative way of studying the leading-order partition functions and consequently leads to a Cardy-like formula for 3d $\mathcal{N}=2$ SCFTs.

\subsection{Supersymmetric Background,  Couplings, and Correlators}

As discussed in \cite{Closset:2012vg},  a global $U(1)$ flavor current $j_\mu$ can be embedded in a real linear superfield $\mathcal{J}$ satisfying $D^2 \mathcal{J} = \overline{D}^2 \mathcal{J} = 0$.  The superfield $\mathcal{J}$ has the following expansion in components:
\be
  \mathcal{J} = J + i \theta j + i \bar{\theta}\, \bar{j} + i \theta\, \bar{\theta}\, K - (\theta \gamma^\mu \bar{\theta})\, j_\mu + \cdots\, .
\ee
The linear superfield $\mathcal{J}$ can couple to a vector superfield $\mathcal{V}$,  which has the following expansion in components:
\be
  \mathcal{V} = \cdots + (\theta \gamma^\mu \bar{\theta})\, a_\mu - i \theta\, \bar{\theta}\, \sigma - i \theta^2\, \bar{\theta}\, \bar{\lambda} + i \bar{\theta}^2\, \theta\, \lambda - \frac{1}{2} \theta^2\, \bar{\theta}^2\, D\, .
\ee
When the superfields $\mathcal{J}$ and $\mathcal{V}$ are coupled together,  {\it i.e.},
\be\label{eq:Coupling J and V}
  2\, \int d^3 x\, d^4 \theta\, \mathcal{J}\, \mathcal{V} = \int d^3 x\, \Big[J D + j_\mu\, a^\mu + K \sigma + \left(\textrm{fermions}\right) \Big]\, ,
\ee
they play the role of a source term in the generating functional:
\be
  e^{- F [\mathcal{V}]} = \Gamma [\mathcal{V}] = \int \mathcal{D} \Phi\, e^{-S[\Phi] - 2 \int d^3 x\, \int d^4 \theta \mathcal{J} \mathcal{V}}\, .
\ee
Hence,
\begin{align}
\begin{split}\label{eq:FirstDerivativesF}
  \frac{\partial F}{\partial D} \Bigg|_{\mathcal{V} = 0} & = - \frac{1}{\Gamma} \frac{\partial \Gamma}{\partial D} \Bigg|_{\mathcal{V} = 0} = - \frac{1}{Z} \Big\langle \int d^3 x\, J \Big\rangle\, ,\\
  \frac{\partial F}{\partial \sigma} \Bigg|_{\mathcal{V} = 0} & = - \frac{1}{\Gamma} \frac{\partial \Gamma}{\partial \sigma} \Bigg|_{\mathcal{V} = 0} = - \frac{1}{Z} \Big\langle \int d^3 x\, K \Big\rangle\, .
\end{split}
\end{align}
Assuming that the correlation functions are normalized,  we obtain
\begin{align}
\begin{split}\label{eq:SecondDerivativesF}
  \frac{\partial^2 F}{\partial D^2} \Bigg|_{\mathcal{V} = 0} & = \int d^3 x\, \int d^3 y\, \langle J(x)\, J(y)\rangle\, ,\\
  \frac{\partial^2 F}{\partial \sigma^2} \Bigg|_{\mathcal{V} = 0} & = \int d^3 x\, \int d^3 y\, \langle K(x)\, K(y)\rangle\, .
\end{split}
\end{align}

In contrast to the flavor symmetries,  the current $j_\mu^{(R)}$ of the $U(1)_R$ symmetry couples to vector in a different way.  It can only be embedded in a supercurrent field $\mathcal{R}_\mu$:
\be
  \mathcal{R}_\mu = j_\mu^{(R)} - i \theta S_\mu - i \bar{\theta}\, \bar{S}_\mu - (\theta \gamma^\nu \bar{\theta})\, \left(2 T_{\mu\nu} + i \epsilon_{\mu\nu\rho}\, \partial^\rho J^{(Z)} \right) - i \theta\, \bar{\theta}\, \left(2 j_\mu^{(Z)} + i \epsilon_{\mu\nu\rho} \partial^\nu j^{(R) \rho} \right) + \cdots
\ee
The supercurrent field $\mathcal{R}_\mu$ can couple to a linearized metric superfield $\mathcal{H}_\mu$,  which in the Wess-Zumino gauge can be expressed as
\be
  \mathcal{H}_\mu = \frac{1}{2} (\theta \gamma^\nu \bar{\theta}) (h_{\mu\nu} - i B_{\mu\nu}) - \frac{1}{2} \theta\, \bar{\theta}\, C_\mu - \frac{i}{2} \theta^2 \bar{\theta}\, \bar{\psi}_\mu + \frac{i}{2} \bar{\theta}^2 \theta \psi_\mu + \frac{1}{2} \theta^2\, \bar{\theta}^2 (A_\mu - V_\mu)\, .
\ee
When the superfields $\mathcal{R}_\mu$ and $\mathcal{H}_\mu$ coupled together,  {\it i.e.},
\be
  2 \int d^3 x\, d^4 \theta\, \mathcal{R}_\mu\, \mathcal{H}^\mu = \int d^3 x \Big[T_{\mu\nu} h^{\mu\nu} - j_\mu^{(R)} \left(A^\mu - \frac{3}{2} V^\mu \right) + i\, j_\mu^{(Z)} C^\mu - J^{(Z)}\, H + (\textrm{fermions}) \Big]\, ,
\ee
these source terms can be added to the action.

For the Abelian currents $j_a^\mu$ at separated points,  including both the flavor $U(1)$ current and the $U(1)_R$ current,  the two-point correlation function of currents have the general expression
\be
  \langle j_a^\mu (x)\, j_b^\nu (y) \rangle = \frac{\tau_{ab}}{16 \pi^2} \left(\delta^{\mu\nu}\, \partial^2 - \partial^\mu \partial^\nu \right)\, \frac{1}{|x - y|^2} + \frac{i\, \kappa_{ab}}{2 \pi}\, \epsilon_{\mu\nu\rho}\, \partial^\rho\, \delta^{(3)} (x - y)\, ,
\ee
where $\tau_{ab}$ for unitary theories is a positive definite matrix,  and the second term ($\sim \kappa_{ab}$) is a contact term caused by the background fields.  Notice the difference in the normalization conventions in \cite{Closset:2012vg} and \cite{Barnes:2005bw} \footnote{In \cite{Barnes:2005bw},  in the absence of the contact term,  the $d$-dimensional two-point correlation function of currents is defined as
\be
  \langle j^\mu_a (x)\, j^\nu_b (y) \rangle = \frac{\tau_{ab}}{(2 \pi)^d} \left(\delta^{\mu\nu} \partial^2 - \partial^\mu \partial^\nu \right)\, \frac{1}{(x - y)^{2 (d-2)}}\, .
\ee
Hence,  for the 3d SCFT the normalizations of $\tau_{ab}$ in \cite{Closset:2012vg} and \cite{Barnes:2005bw} differ by a factor $2/\pi$.
}. In this paper,  we follow the convention of \cite{Closset:2012vg}.

In the 3d Euclidean flat space $\mathbb{R}^3$,  the flavor currents have the correlation functions:
\begin{align}
\begin{split}\label{eq:CurrentsCorrelationsFlat}
  \langle J(x)\, J(y) \rangle & = \frac{\tau_{ff}}{16 \pi^2}\, \frac{1}{|x - y|^2}\, ,\\
  \langle K(x)\, K(y) \rangle & = \frac{\tau_{ff}}{8 \pi^2}\, \frac{1}{|x - y|^4}\, ,\\
  \langle J(x)\, K(y) \rangle & = \frac{\kappa_{ff}}{2 \pi}\, \delta^{(3)} (x - y)\, ,
\end{split}
\end{align}
where for a unitary theory $\tau_{ff}$ is a positive constant,  and $\langle J(x)\, K(y) \rangle$ is a contact term. When $\mathbb{R}^3$ is conformally mapped to $\mathcal{M}_3$,  the above correlation functions of currents should transform correspondingly.  More precisely,  on $\mathcal{M}_3$ we have
\begin{align}
\begin{split}\label{eq:CurrentsCorrelationsCurved}
  \langle J(x)\, J(y) \rangle & = \frac{\tau_{ff}}{16 \pi^2}\, \frac{1}{s(x, y)^2}\, ,\\
  \langle K(x)\, K(y) \rangle & = \frac{\tau_{ff}}{8 \pi^2}\, \frac{1}{s(x, y)^4}\, ,\\
  \langle J(x)\, K(y) \rangle & = \frac{\kappa_{ff}}{2 \pi}\, \delta^{(3)} (x - y)\, \frac{1}{\sqrt{g}}\, ,
\end{split}
\end{align}
with
\be
  s (x,  y) = \Xi^{\frac{1}{2}} (x)\, \Xi^{\frac{1}{2}} (y)\, |x - y|\, ,
\ee
where $\Xi$ is the conformal factor.  Combining \eqref{eq:SecondDerivativesF} and \eqref{eq:CurrentsCorrelationsCurved},  we obtain
\begin{align}
\begin{split}\label{eq:IntegratedCorrelators}
  \frac{\partial^2 F}{\partial D^2} \Bigg|_{\mathcal{V} = 0} & = \frac{\tau_{ff}}{16 \pi^2} \int_{\mathcal{M}_3} d^3 x\, \sqrt{g}\, \int_{\mathcal{M}_3}\, d^3 y\, \sqrt{g}\, \frac{1}{s (x,  y)^2}\, ,\\
  \frac{\partial^2 F}{\partial \sigma^2} \Bigg|_{\mathcal{V} = 0} & = \frac{\tau_{ff}}{8 \pi^2} \int_{\mathcal{M}_3} d^3 x\, \sqrt{g}\, \int_{\mathcal{M}_3}\, d^3 y\, \sqrt{g}\, \frac{1}{s (x,  y)^4}\, ,\\
  \frac{\partial^2 F}{\partial D\, \partial \sigma} \Bigg|_{\mathcal{V} = 0} & = \frac{\kappa_{ff}}{2 \pi} \int_{\mathcal{M}_3} d^3 x\, \sqrt{g}\, .
\end{split}
\end{align}
For different conformally flat manifolds $\mathcal{M}_3$,  the integrals above will provide different constants.

We have discussed above that the linear superfield $\mathcal{J}$ of a $U(1)$ flavor current can be coupled to a vector superfield $\mathcal{V}$.  As shown in \cite{Closset:2012ru},  this vector superfield can be the $U(1)_R$ vector superfield as a special realization.  This particular coupling adds a supersymmetric term to the action:
\be
  \frac{1}{4} \int d^3 x \sqrt{g}\, \mathcal{L}_\mathcal{J} =  \int d^3 x \sqrt{g}\, \left[ j_\mu \frac{1}{4} \left(A^\mu - \frac{1}{2} V^\mu \right) + \frac{1}{4} H K - \frac{1}{16} \left(R - 2 V^\mu V_\mu + 2 H^2 \right) J \right]\, ,
\ee
where an extra factor $\frac{1}{4}$ has been introduced.  In this expression,  $\frac{1}{4} \left(A^\mu - \frac{1}{2} V^\mu\right)$,  $\frac{1}{4} H$ and $- \frac{1}{16} \left(R - 2 V^\mu V_\mu + 2 H^2 \right)$ play the roles of $a^\mu$,  $\sigma$ and $D$ in \eqref{eq:Coupling J and V} respectively.  In our gauge choice \eqref{eq:FixGauge},  we find the classical values of these $U(1)_R$ background fields considered in this paper:
\begin{align}
\begin{split}\label{eq:Bgd Fields}
  a_{1,  R} & = \frac{1}{4} \left(A_1 - \frac{1}{2} V_1\right) = 0\, ,\\
  \sigma_R & = \frac{1}{4} H = \Phi - \frac{1}{4} \Omega = \frac{\pi i}{4 \beta}\, ,\\
  D_R & = - \frac{1}{16} \left(R - 2 V^\mu V_\mu + 2 H^2 \right) = - \frac{1}{16} R + \frac{\pi^2}{4 \beta^2}\, .
\end{split}
\end{align}

\subsection{Review of Results on $S^3$}

Let us first review the results on $S^3$,  originally reported in \cite{Closset:2012vg}.  The $\mathcal{N}=2$ supersymmetry on $S^3$ requires that the background vector multiplet $\mathcal{V}$ has the components
\be
  \sigma = m\, ,\quad D = \frac{i m}{\widetilde{R}}\, ,
\ee
where $\widetilde{R}$ denotes the radius of $S^3$.  In the IR,  the R-symmetry is mixed with the flavors symmetries,
\be
  R(t) = R_0 + \sum_a t^a Q_a\, ,
\ee
and the genuine R-symmetry is given by $t_*$,  which locally maximizes $\mathrm{Re}\, F(t)$.  Hence,  $F$ is a holomorphic function of $(m + i t_* / \widetilde{R})$.  We can expand $F [m + i t_* / \widetilde{R}]$ for small $m$:
\be\label{eq:ExpansionF S3}
  F \left[m + \frac{i t_*}{\widetilde{R}} \right] = F_0 + m \widetilde{R}\, F_1 + \frac{1}{2} (m \widetilde{R})^2\, F_2 + \cdots\, ,
\ee
where a length scale $\widetilde{R}$ has been introduced to make $m \widetilde{R}$ dimensionless.  For $S^3$,  the natural choice of $\widetilde{R}$ is the radius of $S^3$.  In the expansion \eqref{eq:ExpansionF S3},  the term $\sim F_1$ breaks conformal symmetry,  and it can be set to zero by adding a proper counter-term.  Moreover,
\begin{align}
  F_2 & = \frac{1}{\widetilde{R}^2} \frac{\partial^2 F}{\partial m^2}\Bigg|_{m = 0,\, t = t_*}\, ,\\
  \Rightarrow \qquad \mathrm{Re}\, F_2 + i\, \mathrm{Im}\, F_2 & = \frac{1}{\widetilde{R}^2} \frac{\partial^2 F}{\partial m^2}\Bigg|_{m = 0,\, t = t_*} \nonumber\\
  = \frac{1}{\widetilde{R}^2} \frac{\partial^2 F}{\partial \sigma^2}\Bigg|_{m = 0,\, t = t_*} & - \frac{1}{\widetilde{R}^4} \frac{\partial^2 F}{\partial D^2}\Bigg|_{m = 0,\, t = t_*} + \frac{2 i}{\widetilde{R}^3} \frac{\partial^2 F}{\partial \sigma\, \partial D}\Bigg|_{m = 0,\, t = t_*} \, .\label{eq:F2 in IntegratedCorrelators}
\end{align}
Therefore,  the computation of $F_2$ becomes the evaluation of the integrals in \eqref{eq:IntegratedCorrelators} For instance,  for $\mathcal{M}_3 = S^3$ we obtain
\be\label{eq:S3 Re and Im F}
  \mathrm{Re}\, F_2 = \frac{\pi^2}{2}\, \tau_{ff}\, ,\quad \mathrm{Im}\, F_2 = - 2 \pi\, \kappa_{ff}\, .
\ee
For the 3d SCFT without extra flavor symmetries,  we can view the R-symmetry as a flavor symmetry with $t_* = 0$.  Then,  the results discussed above for a flavor symmetry still hold for this case,  with $\tau_{ff}$ replaced by $\tau_{RR}$.

\subsection{Results on $S^1 \times S^2$}\label{sec:S1S2 F analysis}

The 3d space $S^1 \times S^2$ can be conformally mapped to a 3d Euclidean space.  This can be done in two steps.  Starting from the metric \eqref{eq:metric-3} of $S^1 \times S^2$:
\be
  ds^2 = d\tau^2 + L^2 \Big(d\theta^2 + \mathrm{sin}^2 \theta\, d\widetilde{\varphi}^2\Big)\, ,
\ee
where we have explicitly introduced the radius $L$ for $S^2$.  Next,  we can define a new coordinate
\be
  d\tau = \frac{L\, d\psi}{\mathrm{sin} \psi}\quad\Rightarrow\quad e^{\tau / L} = \mathrm{tan}\, \frac{\psi}{2}\, .
\ee
This definition also indicates the range of $\psi$ is given by
\be
  \tau \in [0,\, \beta]\quad\Rightarrow\quad \mathrm{tan}\, \frac{\psi}{2} \in [1,\, e^{\beta / L}]\, .
\ee
In the new coordinates $(\psi, \, \theta,\, \widetilde{\varphi})$,  the metric of $S^1 \times S^2$ can be conformally mapped into the metric of $S^3$:
\be\label{eq:metric-4}
  ds^2 = \frac{L^2}{\mathrm{sin}^2 \psi} \Big[d\psi^2 + \mathrm{sin}^2 \psi \left(d\theta^2 + \mathrm{sin}^2 \theta\, d\widetilde{\varphi}^2 \right) \Big]\, .
\ee
Next,  we perform a stereographic projection to conformally map $S^3$ into $\mathbb{R}^3$:
\be
  ds^2 = \frac{1}{\mathrm{sin}^2 \psi}\cdot \frac{4\, L^4}{(L^2 + x_i^2)^2}\, ds^2_{\text{flat}} = \frac{1}{\mathrm{sin}^2 \psi}\cdot \frac{4\, L^4}{(L^2 + x_i^2)^2}\, \sum_{i=1}^3 dx_i^2\, ,
\ee
where
\be\label{eq:ChangeCoord}
  x_i^2 = L^2\, \mathrm{cot}^2 \left(\frac{\psi}{2} \right)\quad\Rightarrow\quad |x| = \frac{L}{\mathrm{tan}\, \frac{\psi}{2}}\in [L\, e^{-\beta / L},\, L]\, .
\ee
The range of $|x|$ indicates that this metric covers only a shell region of $\mathbb{R}^3$.  Using the relation \eqref{eq:ChangeCoord} between coordinates,  we find that the metric \eqref{eq:metric-4} takes a simple expression
\be\label{eq:metric-5}
  ds^2 = \frac{L^2}{|x|^2}\, ds^2_{\text{flat}}\, .
\ee

Up to now,  we have only rewritten the metric \eqref{eq:metric-3} of $S^1 \times S^2$ in different coordinates.  In this paper,  we also consider the background $S^1 \times \textrm{(rotating  $S^2$)}$.  For this background with rotating $S^2$,  a new scale $\widetilde{\beta} \equiv \beta - 2 \pi i / \Omega$ emerges,  which is more relevant in the Kaluza-Klein reduction of $S^1$,  or more precisely,  in the Cardy-like limit
\be\label{eq:CardyLimit}
  \beta \to 0\, ,\quad \text{for the non-rotating case};\quad \omega \sim \beta \to 0\, ,\quad \text{for the rotating case},
\ee
where $\omega \equiv 2\, \Omega\, \beta$ with $\Omega \sim \mathcal{O} (1)$.  Therefore,  in the Cardy-like limit \eqref{eq:CardyLimit},
\be
  \widetilde{\beta} \equiv \beta - \frac{2 \pi i}{\Omega} \sim \mathcal{O} (1) \quad\Rightarrow\quad \widetilde{\beta} \gg \beta\, .
\ee
For $S^1 \times \textrm{(rotating  $S^2$)}$ we will consider the following metric
\be\label{eq:metric-6}
  ds^2 = \frac{\widetilde{\beta}^2}{|x|^2}\, ds^2_{\text{flat}} = \Xi^2\, \sum_{i=1}^3 dx_i^2\, ,
\ee
which is conformally equivalent to \eqref{eq:metric-5},  hence also conformally flat but with a different conformal factor
\be
  \Xi \equiv \frac{\widetilde{\beta}}{|x|}\, .
\ee
In accordance with \eqref{eq:metric-6},  the scale $L$ in \eqref{eq:metric-4} and \eqref{eq:ChangeCoord} should also be replaced by the new scale $\widetilde{\beta}$.  Hence,  the range of $|x|$ now becomes
\be\label{eq:ChangeCoord New}
  |x| \in \big[\widetilde{\beta}\, e^{-\beta / \widetilde{\beta}},\, \widetilde{\beta}\big]\, .
\ee
Similar to the relation \eqref{eq:ExpansionF S3} for $S^3$,  we can expand the free energy $F$ in $m$ on $S^1 \times \textrm{(rotating  $S^2$)}$,  or equivalently,  $S^1 \times S^2$ with double periods characterized by $\widetilde{\beta}$. 
\be\label{eq:ExpansionF conformal S3}
  F \left[m + \frac{i t_*}{\widetilde{\beta}} \right] = F_0 + m \widetilde{\beta}\, F_1 + \frac{1}{2} (m \widetilde{\beta})^2\, F_2 + \cdots\, .
\ee
The only difference bewteen \eqref{eq:ExpansionF conformal S3} and \eqref{eq:ExpansionF S3} is that we use the length scale $\widetilde{\beta}$ to replace $R$.  In both expressions,  $m$ is the classical value of $\sigma$,  the scalar component of the vector multiplet coupled to the component $K$ of the flavor current.

The background fields coupled to the flavor current cannot take the values of the $U(1)_R$ vector multiplet given by \eqref{eq:Bgd Fields},  which are inconsistent with the Cardy-like limit \eqref{eq:CardyLimit}.  Instead,  we find the consistent values of the background fields $\sigma$ and $D$ in the Cardy-like limit to be
\be
  \sigma = m\, ,\quad D = - \frac{\pi i}{\widetilde{\beta}} m \, .
\ee
Using these relations,  $F_2$ can be computed from the expansion \eqref{eq:ExpansionF conformal S3} as
\begin{align}
  F_2 & = \frac{1}{\widetilde{\beta}^2} \frac{\partial^2 F}{\partial m^2}\Bigg|_{m = 0,\, t = t_*} \nonumber\\
  {} & = \frac{1}{\widetilde{\beta}^2} \frac{\partial^2 F}{\partial \sigma^2}\Bigg|_{m = 0,\, t = t_*} - \frac{\pi^2}{\widetilde{\beta}^4} \frac{\partial^2 F}{\partial D^2}\Bigg|_{m = 0,\, t = t_*} - \frac{2 \pi i}{\widetilde{\beta}^3} \frac{\partial^2 F}{\partial \sigma\, \partial D}\Bigg|_{m = 0,\, t = t_*}\, , \label{eq:F2 in IntegratedCorrelators for S1 S2}
\end{align}
which eventually turns out to be independent of the length scale $\widetilde{\beta}$.

To determine $F_2$ we should first evaluate the integrated correlators \eqref{eq:IntegratedCorrelators} for $\mathcal{M}_3 = S^1 \times \textrm{(rotating  $S^2$)}$,  or equivalently,  $S^1 \times S^2$ with double periods characterized by $\widetilde{\beta}$,  and then compute $F_2$ using \eqref{eq:F2 in IntegratedCorrelators for S1 S2}.  We choose the non-rotating $S^1 \times S^2$ description.  From \eqref{eq:metric-6},  we have seen that $S^1 \times S^2$ can be conformally mapped into an Euclidean flat space $\mathbb{R}^3$.  Hence,  for $\mathcal{M}_3 = S^1 \times S^2$ the integrand \eqref{eq:IntegratedCorrelators} becomes
\be
  \frac{1}{s(x,\, y)^2} = \frac{1}{\Xi_x} \cdot \frac{1}{\Xi_y} \cdot \frac{1}{|x - y|^2}\, .
\ee
Consequently,
\begin{align}
  \frac{\partial^2 F}{\partial D^2} \Bigg|_{\mathcal{V} = 0} & = \frac{\tau_{ff}}{16 \pi^2} \int_{S^1 \times S^2} d^3 x\, \sqrt{g}\, \int_{S^1 \times S^2}\, d^3 y\, \sqrt{g}\, \frac{1}{s (x,  y)^2} \nonumber\\
  {} & = \frac{\tau_{ff}}{16 \pi^2} \int d^3 x \left(\frac{\widetilde{\beta}}{|x|} \right)^3 \int d^3 y \left(\frac{\widetilde{\beta}}{|y|} \right)^3\, \frac{|x|\, |y|}{\widetilde{\beta}^2}\cdot \frac{1}{|x - y|^2} \nonumber\\
  {} & = \frac{\tau_{ff}}{16 \pi^2}\, \widetilde{\beta}^4 \int 4 \pi r_x^2 dr_x \int 2 \pi r_y^2 dr_y\, \mathrm{sin}\, \theta_y\, d\theta_y \frac{1}{r_x^2\, r_y^2\, \left(r_x^2 + r_y^2 - 2\, r_x\, r_y\, \mathrm{cos}\, \theta_y \right)} \nonumber\\
  {} & = \frac{\tau_{ff}}{2}\, \widetilde{\beta}^4 \int_{e^{-\beta / \widetilde{\beta}}}^1 dr_x \int_{e^{-\beta / \widetilde{\beta}}}^1 dr_y \frac{\mathrm{log} \Big[ (r_x + r_y)^2\Big] - \mathrm{log} \Big[(r_x - r_y)^2 \Big]}{2\, r_x\, r_y}\, ,\label{eq:d2F/dD2}
\end{align}
where we have defined $r_x \equiv |x| / \widetilde{\beta}$ and $r_y \equiv |y| / \widetilde{\beta}$,  which have the range $[e^{-\beta / \widetilde{\beta}},\, 1]$ according to \eqref{eq:ChangeCoord New}.  In practice,  we first evaluate the integral over $r_x$,  expand the result in $\beta / \widetilde{\beta}$,  and then evaluate the integral over $r_y$.  It can be shown that the final result of \eqref{eq:d2F/dD2} is of the order $\mathcal{O} \left( (\beta / \widetilde{\beta})^2 \right)$.  Similarly,  we can evalute:
\begin{align}
  \frac{\partial^2 F}{\partial \sigma^2} \Bigg|_{\mathcal{V} = 0} & = \frac{\tau_{ff}}{8 \pi^2} \int_{\mathcal{M}_3} d^3 x\, \sqrt{g}\, \int_{\mathcal{M}_3}\, d^3 y\, \sqrt{g}\, \frac{1}{s (x,  y)^4} \nonumber\\
  {} & = \frac{\tau_{ff}}{8 \pi^2} \int d^3 x \left(\frac{\widetilde{\beta}}{|x|} \right)^3 \int d^3 y \left(\frac{\widetilde{\beta}}{|y|} \right)^3\, \frac{|x|^2\, |y|^2}{\widetilde{\beta}^4}\cdot \frac{1}{|x - y|^4} \nonumber\\
  {} & = \frac{\tau_{ff}}{8 \pi^2}\, \widetilde{\beta}^2 \int 4 \pi r_x^2\, dr_x \int 2 \pi r_y^2 dr_y\, \mathrm{sin}\, \theta_y\, d\theta_y \frac{1}{r_x\, r_y\, \left(r_x^2 + r_y^2 - 2\, r_x\, r_y\, \mathrm{cos}\, \theta_y \right)^2} \nonumber\\
  {} & = \tau_{ff}\, \widetilde{\beta}^2 \int_{e^{-\beta / \widetilde{\beta}}}^1 dr_x \int_{e^{-\beta / \widetilde{\beta}} - \epsilon}^{1 + \epsilon} dr_y\, \frac{2\, r_x\, r_y}{(r_x^2 - r_y^2)^2}\, ,
\end{align}
where we introduced a small parameter $\epsilon$ to regularize the integral.  More precisely,  we choose $\epsilon = \beta$ to prevent some logarithmic divergences.  The result after regularization is
\be
  \frac{\partial^2 F}{\partial \sigma^2} \Bigg|_{\mathcal{V} = 0} = - \tau_{ff}\, \widetilde{\beta}^2 + \mathcal{O} (\beta / \widetilde{\beta})\, .
\ee
Furthermore,
\be
  \frac{\partial^2 F}{\partial D\, \partial \sigma} \Bigg|_{\mathcal{V} = 0} = \frac{\kappa_{ff}}{2 \pi} \int_{\mathcal{M}_3} d^3 x\, \sqrt{g} = \frac{\kappa_{ff}}{2 \pi} \int d^3 x \left(\frac{\widetilde{\beta}}{|x|} \right)^3 = \frac{\kappa_{ff}}{2 \pi}\, \widetilde{\beta}^3 \int_{e^{-\beta / \widetilde{\beta}}}^1 4 \pi \frac{dr_x}{r_x} = 2\, \kappa_{ff}\, \widetilde{\beta}^2\, \beta\, .
\ee
Therefore,  using \eqref{eq:F2 in IntegratedCorrelators for S1 S2} we obtain
\begin{align}
  F_2 & = \frac{1}{\widetilde{\beta}^2} \frac{\partial^2 F}{\partial m^2}\Bigg|_{m = 0,\, t = t_*} = \frac{1}{\widetilde{\beta}^2} \frac{\partial^2 F}{\partial \sigma^2}\Bigg|_{m = 0,\, t = t_*} - \frac{\pi^2}{\widetilde{\beta}^4} \frac{\partial^2 F}{\partial D^2}\Bigg|_{m = 0,\, t = t_*} - \frac{2 \pi i}{\widetilde{\beta}^3} \frac{\partial^2 F}{\partial \sigma\, \partial D}\Bigg|_{m = 0,\, t = t_*}\nonumber\\
  {} & = - \tau_{ff} + \mathcal{O} (\beta / \widetilde{\beta})\, .
\end{align}

Knowing the value of $F_2$ in this case,  we can continue expanding the free energy $F$ in \eqref{eq:ExpansionF conformal S3} as
\begin{align}
  F & = F_0 + m \widetilde{\beta}\, F_1 + \frac{1}{2} (m \widetilde{\beta})^2\, F_2 + \cdots \nonumber\\
  {} & = F_0 + m \left(\beta - \frac{2 \pi i}{\Omega} \right) F_1 + \frac{1}{2} m^2 \left(\beta - \frac{2 \pi i}{\Omega} \right)^2\, F_2 + \cdots\, .\label{eq:ExpansionF S1S2}
\end{align}
The aim is to extract the dominant contribution to $F$ \eqref{eq:ExpansionF S1S2} in the Cardy-like limit \eqref{eq:CardyLimit},  i.e.,  $\omega \equiv 2\, \Omega\, \beta \sim \beta \to 0$.  Let us analyze the expansion \eqref{eq:ExpansionF S1S2} term by term.  The term $F_1$ breaks conformal symmetry,  and can be removed by adding a counter-term.  Because in the Cardy-like limit \eqref{eq:CardyLimit} the $S^1 \times \textrm{(rotating  $S^2$)}$ essentially becomes $S^2$,   the term $F_0$ will take the form like in a 2d CFT:
\be
  F_0 \sim c\, \textrm{log} (\Lambda r_0) + \cdots\, ,
\ee
which in the large-$N$ limit is of the order $G_2^{-1} \sim \beta G_3^{-1} \sim \beta N^{3/2} \text{ or }  \beta N^{5/3}$ and hence subleading compared to the $F_2$ term of the order $\mathcal{O} (N^{3/2})$.  Therefore,  we only need to focus on the $F_2$ term.  Taking into account the constraint \eqref{eq:constraint from TTI} in the Cardy-like limit \eqref{eq:CardyLimit},  we should require the scaling
\be
  \widetilde{\Delta} \equiv 2 m \beta \sim \mathcal{O} (1)\, ,\quad\Rightarrow\quad m \sim \mathcal{O} (\beta^{-1})\, .
\ee
These scalings have been seen from the Cardy limit on the gravity side \cite{David:2020ems}:
\be
  \omega \sim \epsilon\, ,\quad \widetilde{\Delta} \sim \mathcal{O} (1)\, .
\ee
At the quadratic order in $m \widetilde{\beta}$,
\be\label{eq:F}
  F \supset \frac{1}{2} (m \widetilde{\beta})^2\, F_2 = \frac{1}{2} m^2 \left(\beta - \frac{2 \pi i}{\Omega} \right)^2 F_2 = \frac{1}{2} m^2 \left[\beta^2 - \beta \frac{4 \pi i}{\Omega} + \left(- \frac{2 \pi i}{\Omega} \right)^2 \right] F_2\, .
\ee
The first term proportional to $m^2\beta^2$  is order ${\cal O}(\beta^0)$ in the scaling we consider.  The second term is our main focus, and it is of order  ${\cal O}(\beta^{-1})$ in the scaling limit we are interested in. Note that the last term above is order ${\cal O}(\beta^{-2})$ and can be renormalized by a cosmological constant counter-term \cite{Closset:2012vg} in the Cardy limit $\beta \to 0$:
\be
  \Lambda^2 \int_{S^2} d^2 x\, \sqrt{g}\, .
\ee
Consequently,  the important term in \eqref{eq:F} is
\be
  F \supset \frac{1}{2} m^2 \left(- \beta \frac{4 \pi i}{\Omega} \right) F_2 \sim \mathcal{O} (\beta^{-1})\, ,
\ee
which seems divergent in the limit $\beta \to 0$.  However,  this is expected,  and the reason is the following. 
For the BPS case,  the free energy $F$ in the BPS limit can be defined as $dF (T)/dT\big|_{T=0}$.  Hence,  the free energy is related to the AdS black hole entropy function,  i.e.,
\be
  F = - S \sim \mathcal{O} (\beta^{-1})\, ,
\ee
which is the same as the result from the gravitational Cardy limit \cite{David:2020ems}:
\be
  S \sim \frac{1}{\epsilon}\, .
\ee
In summary,  after dropping the vanishing term and removing the divergent term by adding appropriate counter-terms,  the leading contribution to $F$ in the Cardy-like limit \eqref{eq:CardyLimit} comes from the $F_2$ term
\begin{align}
  F & = \frac{1}{2} m^2 \beta \left(- \frac{4 \pi i}{\Omega}\right) F_2 + \cdots = - 2 \pi i \frac{(m \beta)^2}{\Omega \beta} F_2 + \cdots = 2 \pi i \frac{(m \beta)^2}{\Omega \beta}\, \tau_{ff} + \cdots\, ,
\end{align}
where the dots include subleading terms in $\beta / \widetilde{\beta}$.  Again,  $m$ is the classical value of the scalar $\sigma$ in the background vector multiplet,  which is coupled to the flavor current multiplet.  We will justify shortly that $m$ is proportional to the chemical potential $\Phi$ of the electric charge.

\subsection{Examples}

In this subsection,  we consider several explicit examples, previously discussed in Sec.~\ref{sec:2D} from the 2d perspective.   From the 3d point of view, we will show that in every case, $\tau_{RR}$ yields the leading-order scaling in $N$ of the free energy $F$.

\subsubsection{The ABJM Theory on $S^1 \times \textrm{(rotating  $S^2$)}$}

Let us take the 3d ABJM theory as an example.  As  discussed in the previous subsection,  for the 3d SCFT without flavor symmetries such as the ABJM theory,  we can apply a trick of treating the R-symmetry as a ``probe'' flavor symmetry.  Consequently,  the discussions above for $U(1)$ flavor symmetries maintain for the R-symmetry,  with $\tau_{ff}$ replaced by $\tau_{RR}$.  For 3d $\mathcal{N}=2$ SCFT with holographic dual,  $\tau_{RR}$ can be computed as follows \cite{Barnes:2005bw}:
\be
  \tau_{RR} = \frac{8 \pi N^{3/2}}{3 \sqrt{6\, \textrm{Vol} (Y_7)}}\, ,
\ee
where a normalization factor $2 / \pi$ has been introduced due to the different normalization conventions of $\tau_{IJ}$ in \cite{Closset:2012vg} and \cite{Barnes:2005bw},   and $\textrm{Vol} (Y_7) = \pi^4 / 3 k$ for $Y_7 = S^7 / \mathbb{Z}_k$.  Hence,  for the 3d ABJM theory,  $\tau_{RR}$ has the value:
\be
  \tau_{RR} = \frac{4 \sqrt{2}\, k^{\frac{1}{2}}\, N^{\frac{3}{2}}}{3 \pi}\, .
\ee
Therefore,  in the Cardy-like limit \eqref{eq:CardyLimit} the renormalized free energy of the ABJM theory on $S^1 \times \textrm{(rotating  $S^2$)}$ is
\begin{align}
  F_{S^1 \times (\text{rot.} S^2)}^{\textrm{ABJM}} & = 2 \pi i \frac{(m \beta)^2}{\Omega \beta}\, \tau_{RR} + \cdots = 2 \pi i \frac{(m \beta)^2}{\Omega \beta}\, \frac{4 \sqrt{2}\, k^{\frac{1}{2}}\, N^{\frac{3}{2}}}{3 \pi} + \cdots \nonumber\\
  {} & = \frac{4 \sqrt{2}\, i\, k^{\frac{1}{2}}\, N^{\frac{3}{2}}\, \widetilde{\Delta}^2}{3\, \widetilde{\omega}} + \cdots\, ,
\end{align}
where we have defined $\widetilde{\Delta} \equiv 2 m \beta$ and $\widetilde{\omega} \equiv 2 \Omega \beta$,  which are consistent with the definitions in \cite{Nian:2019pxj}.  From these definitions we see that $m$ is proportional to the chemical potential $\Phi$.

This result of the ABJM theory can be used to study the asymptotically AdS$_4$ black hole via the AdS/CFT correspondence.  At the leading order,  the entropy function of the electrically charged AdS$_4$ BPS black hole is given by the Legendre transform of $\textrm{log}\, Z = - F$ subject to a constraint,  {\it i.e.},
\be\label{eq:EntropyFct}
  S (\widetilde{\Delta},\, \widetilde{\omega}) = - \frac{4 \sqrt{2}\, i\, k^{\frac{1}{2}}\, N^{\frac{3}{2}}\, \widetilde{\Delta}^2}{3\, \widetilde{\omega}} + \widetilde{\omega} J + 4 \widetilde{\Delta} Q + \Lambda \left(4 \widetilde{\Delta} - \widetilde{\omega} - 2 \pi i \right)\, ,
\ee
where we have introduced a Lagrange multiplier $\Lambda$ to impose the constraint \eqref{eq:Constraint}.  The entropy function \eqref{eq:EntropyFct} is exactly the special case with 4 equal chemical potentials of the generic one in \cite{Nian:2019pxj}.  In fact,  we can obtain the generic entropy function by taking a refined expansion from the beginning.  More precisely,  instead of \eqref{eq:ExpansionF conformal S3} we can expand $F$ as
\begin{align}
  F_{S^1 \times (\text{rot.} S^2)}^{\textrm{ABJM}} \left[m + \frac{i t_*}{\widetilde{\beta}} \right] & = F_0 + \sqrt[4]{(m_1 \widetilde{\beta}) (m_2 \widetilde{\beta}) (m_3 \widetilde{\beta}) (m_4 \widetilde{\beta})}\, F_1 \nonumber\\
  {} & \quad + \frac{1}{2} \sqrt{(m_1 \widetilde{\beta}) (m_2 \widetilde{\beta}) (m_3 \widetilde{\beta}) (m_4 \widetilde{\beta})}\, F_2 + \cdots\, ,
\end{align}
where $m_i$ can have generically different values of the same order.  Then,  following similar steps,  we obtain the generic entropy function for AdS$_4$ BPS black holes
\be\label{eq:generic EntropyFct}
  S (\widetilde{\Delta}_I,\, \widetilde{\omega}) = - \frac{4 \sqrt{2}\, i\, k^{\frac{1}{2}} N^{\frac{3}{2}}}{3} \frac{\sqrt{\widetilde{\Delta}_1 \widetilde{\Delta}_2 \widetilde{\Delta}_3 \widetilde{\Delta}_4}}{\widetilde{\omega}} + \widetilde{\omega} J + \sum_{I=1}^4 \widetilde{\Delta}_I Q_I + \Lambda \left(\sum_{I=1}^4 \widetilde{\Delta}_I - \widetilde{\omega} - 2 \pi i \right)\, ,
\ee
which is exactly the same as the one obtained in \cite{Nian:2019pxj}.

\subsubsection{The Mass Deformed ABJM Theory}

In the above,  we focus on the rotating AdS$_4$ black holes.  In fact,  the formalism can be adapted to the non-rotating AdS$_4$ black holes.  This can be done by replacing $\widetilde{\beta}$ with the period $\beta$ of the $S^1$ circle,  where the radius $L$ of $S^2$ in the metric \eqref{eq:metric} can be effectively set to $1$.

The background scalar has already contained the factors $2 \pi$ and $\widetilde{R}_\beta \equiv \beta / 2 \pi$,  which can be seen from the constraint \eqref{eq:FixGauge} \eqref{eq:Constraint},  the definitions \eqref{eq:def chemical potentials} and $y_i \equiv e^{2 \pi i \nu_i} = e^{i \Delta_i}$ (with trivial holonomies $a_{0a} = a_{0i}^F = 0$).
\be\label{eq:NewConstraint}
  \beta H = \pi i\quad (\textrm{mod } 2 \pi i) \quad\Rightarrow\quad \Delta_ i = 2 \pi \nu_i = 2 \pi i \widetilde{R}_\beta m_i = i \beta m_i = \pi\quad (\textrm{mod } 2 \pi)\, .
\ee

For the zero magnetic flux sector,  the free energy for the ABJM theory with the Chern-Simons level $k=1$ on $S^1 \times S^2$ is
\begin{align}
  F_{S^1 \times S^2}^{\textrm{ABJM}} & = \frac{1}{2} m^2 \beta^2 F_2 + \cdots = - \frac{1}{2} m^2 \beta^2 \tau_{RR} + \cdots \nonumber\\
  {} & \simeq - \frac{1}{2} (m \beta)^2 \frac{4 \sqrt{2}\, N^{\frac{3}{2}}}{3 \pi} = - (m \beta)^2 \frac{2 \sqrt{2}\, N^{\frac{3}{2}}}{3 \pi} = \overline{\Delta}^2 \frac{2 \sqrt{2}\, N^{\frac{3}{2}}}{3 \pi}\, ,
\end{align}
where $\overline{\Delta} \equiv i m \beta$,  and $\tau_{ff}$ is replaced by $\tau_{RR}$ for the ABJM theory on $(S^1 \times S^2) \times Y_7$ with $Y_7 = S^7$.  In the second line above,  we keep only the leading order.

The result above is in the degenerate case.  More generally,
\be
  F_{S^1 \times S^2}^{\textrm{ABJM}} = \sqrt{\Delta_1\, \Delta_2\, \Delta_3\, \Delta_4 }\, \frac{2 \sqrt{2}\, N^{\frac{3}{2}}}{3 \pi}\, ,
\ee
with the new constraint originated from the contraint on the flavor fugacities
\be\label{eq:NewConstraint 1}
  \prod_{I=1}^4 y_I \equiv \prod_{I=1}^4 e^{i\, \Delta_I} = 1\quad\Rightarrow\quad \sum_{I=1}^4 \Delta_I = 2 \pi\, .
\ee

Now,  let us generalize the result above to the mass deformed ABJM (mABJM) theory by restricting the chemical potential $\Delta_1$ with a superpotential \cite{Bobev:2018uxk}.  The free energy becomes
\be\label{eq:F mABJM}
  F_{S^1 \times S^2}^{\textrm{mABJM}} = \sqrt{\Delta_2\, \Delta_3\, \Delta_4 }\, \frac{2 \sqrt{2 \pi}\, N^{\frac{3}{2}}}{3 \pi}\, ,
\ee
with the new constraint
\be
  \sum_{I=2}^4 \Delta_I = \pi\, .
\ee
The expression \eqref{eq:F mABJM} exactly matches the logarithm of the 3d supersymmetric partition function,  {\it i.e.},  $- \textrm{log}\, Z_{S^1 \times S^2}^{\textrm{mABJM}}$,  if $\mathfrak{n}_i$ are chosen to minimize the supersymmetric partition function \eqref{eq:A-twist part fct mABJM}.

\subsubsection{The AdS$_4$ Black Hole in Massive IIA Supergravity}
For the field theory dual to the AdS$_4$ solution in massive IIA supergravity,  the $S^3$ free energy at the leading order $\mathcal{O} (N^{5/3})$ is \cite{Guarino:2015jca,  Fluder:2015eoa}
\be
  F_{S^3}^{\textrm{IIA}} = \frac{3 \sqrt{3} \pi}{20\times 2^{1/3}} \left(1 - \frac{i}{\sqrt{3}} \right) k^{1/3} N^{5/3} \left(2\, \textrm{Tr}\, R^3 \right)^{2/3}\, .
\ee
Suppose that $(m \widetilde{R})^2 = \left(\sqrt{6}\, \textrm{Tr}\, R^3 \right)^{2/3}$.  Using the relation \eqref{eq:S3 Re and Im F},  we can read off the coefficients $\tau_{RR}$ and $\kappa_{RR}$ from $F_{S^3}^{\textrm{IIA}}$:
\be
  \tau_{RR} = \frac{3\times 3^{1/6} k^{1/3} N^{5/3}}{5\pi}\, ,\quad \kappa_{RR} = \frac{3^{2/3} k^{1/3} N^{5/3}}{40}\, ,
\ee
which should remain the same for this theory defined on $S^1 \times S^2$.  Using this value of $\tau_{RR}$,  we can compute at the leading order:
\begin{align}
  \textrm{Re}\left( F_{S^1 \times S^2}^{\textrm{IIA}}\right) & \simeq \frac{1}{2} m^2 \beta^2\, \textrm{Re} \left(F_2\right) \simeq - \frac{1}{2} m^2 \beta^2 \tau_{RR} \equiv \frac{1}{2}\, \overline{\Delta}^2\cdot \tau_{RR} \nonumber\\
  {} & = \frac{1}{2} \left(\textrm{Tr}\, R^3 \right)^{2/3}\cdot \tau_{RR} = \frac{1}{2} \left(\frac{1}{2} \Delta_1\, \Delta_2\, \Delta_3 \right)^{2/3}\cdot \tau_{RR} \nonumber\\
  {} & = \frac{3^{7/6}}{5 \pi \times 2^{5/3}} k^{1/3}\, N^{5/3}\, \left(\Delta_1\, \Delta_2\, \Delta_3 \right)^{2/3}\, ,\label{eq:ReF IIA}
\end{align}
where $\overline{\Delta} \equiv i m \beta = \left(\textrm{Tr}\, R^3 \right)^{1/3}$,  and $\textrm{Tr}\, R^3$ takes the value $\frac{1}{2} \Delta_1 \Delta_2 \Delta_2$ for the AdS$_4$ black hole in massive IIA supergravity \cite{Hosseini:2017fjo,  Benini:2017oxt}.  The chemical potentials $\Delta_I$'s satisfy the constraint originated from the one on the flavor fugacities
\be\label{eq:NewConstraint 1}
  \prod_{I=1}^3 y_I \equiv \prod_{I=1}^3 e^{i\, \Delta_I} = 1\quad\Rightarrow\quad \sum_{I=1}^3 \Delta_I = 2 \pi\, .
\ee
The expression \eqref{eq:ReF IIA} exactly matches the logarithm of the 3d supersymmetric partition function,  {\it i.e.},  $-\, \textrm{Re}\, \textrm{log}\, Z_{S^1 \times S^2}^{\textrm{IIA}}$,  if $\mathfrak{n}_i$ are chosen to minimize the supersymmetric partition function \eqref{eq:A-twist part fct IIA}.

According to the analysis in Sec.~\ref{sec:S1S2 F analysis},  the imaginary part of the free energy subleading compared to the real part in the expansion of $\beta / \widetilde{\beta} = \beta / L$.  Nevertheless,  we can analyze $\textrm{Im}\left( F_{S^1 \times S^2}^{\textrm{IIA}}\right)$ up to a proportionality constant:
\begin{align}
  \textrm{Im}\left( F_{S^1 \times S^2}^{\textrm{IIA}}\right) & \propto - \frac{1}{2} m^2 \beta^2 \kappa_{RR} = \frac{1}{2}\, \overline{\Delta}^2\cdot \kappa_{RR} \nonumber\\
  {} & = \frac{1}{2} \left(\textrm{Tr}\, R^3 \right)^{2/3}\cdot \kappa_{RR} = \frac{1}{2} \left(\frac{1}{2} \Delta_1\, \Delta_2\, \Delta_3 \right)^{2/3}\cdot \kappa_{RR} \nonumber\\
  {} & = \frac{3^{2/3}}{80\times 2^{2/3}}\, k^{1/3}\, N^{5/3}\, \left(\Delta_1\, \Delta_2\, \Delta_3 \right)^{2/3}\, .
\end{align}
This expression is also proportional to $-\, \textrm{Im}\, \textrm{log}\, Z_{S^1 \times S^2}^{\textrm{IIA}}$,  if $\mathfrak{n}_i$ are chosen to minimize the 3d supersymmetric partition function \eqref{eq:A-twist part fct IIA}.

It was shown in \cite{Hosseini:2017fjo} that the topologically twisted index of this theory can also be expressed in terms of the anomaly coefficient $a$ of the 4d ``parent'' $\mathcal{N}=1$ superconformal field theory on $S^2 \times T^2$ and the right-moving central charge $c_r$ of the 2d $\mathcal{N}=(0,2)$ theory on $T^2$:
\be\label{eq:logZ IIA central charges}
  \textrm{log}\, Z_{S^1 \times S^2}^{\textrm{IIA}} (\mathfrak{n}_I,\, \Delta_I) = - \frac{3^{7/6}\, \pi}{5 \times 2^{10/3}} \left(1 - \frac{i}{\sqrt{3}} \right)\, (k N)^{1/3}\, \frac{c_r (\mathfrak{n}_I,\, \Delta_I)}{a (\Delta_I)^{1/3}}\, ,
\ee
with
\begin{align}
\begin{split}
  a (\Delta_I) & = \frac{9}{32}\, \textrm{Tr}\, R^3 (\Delta_I) - \frac{3}{32}\, \textrm{Tr}\, R (\Delta_I)\, ,\\
  c_r (\mathfrak{n}_I,\, \Delta_I) & = 3\, \textrm{Tr}\, R^3 (\Delta_I) + \pi \sum_I \left[\left(\mathfrak{n}_I - \frac{\Delta_I}{\pi} \right)\, \frac{\partial\, \textrm{Tr}\, R^3 (\Delta_I)}{\partial\, \Delta_I} \right]\, .
\end{split}
\end{align}
Hence,  applying the relation between the 3d supersymmetric partition function on the A-twisted background for $S^1\times S^2$ and the topologically twisted index,  $\mathfrak{n}_I = r_I / 2$,  we obtain the following relation between the 2d,  the 3d and the 4d SCFT quantities:
\be\label{eq:relation between c,  a and tau}
  \frac{1}{2} \left(\textrm{Tr}\, R^3 \right)^{2/3}\cdot \tau_{RR}^{\textrm{min}} = - \frac{3^{7/6}\, \pi}{5 \times 2^{10/3}}\, (k N)^{1/3}\, \left[\frac{c_r (\frac{1}{\pi}\, \Delta_I,\, \Delta_I)}{a (\Delta_I)^{1/3}} \right]_{\textrm{min}}\, ,
\ee
which is compatible with the 2d $c$-extremization,  the 3d $\tau_{RR}$-minimization and the 4d $a$-maximization.  As pointed out in \cite{Amariti:2015ybz,  Amariti:2021cpk},  the 3d $\tau_{RR}$-minimization is also related to the 3d $F$-maximization via:
\be
  \tau_{RR} = \frac{4}{\pi^2}\, \frac{\left(F_{S^3}^{\textrm{max}} \right)^3}{\left(F_{S^3} \right)^2} \quad\Rightarrow\quad \tau_{RR}^{\textrm{min}} = \frac{4}{\pi^2}\, F_{S^3}^{\textrm{max}}\, .
\ee
Hence,  we see that the consistency of various extremization principles in different dimensions becomes manifest in this case.

\subsubsection{The Family of Theories with the Scaling $\sim N^{5/3}$} 

For the more general class of theories with the scaling $\sim N^{5/3}$ \cite{Fluder:2015eoa,  Hosseini:2022vho},  their $S^3$ partition functions and consequently the $\tau_{RR}$ and the $\kappa_{RR}$ coefficients have not been worked out in the literature.  However,  based on the discussions above,  we expect that the following relations still hold:
\begin{align}
\begin{split}\label{eq:F 5/3 expected}
  \textrm{Re}\left( F_{S^1 \times S^2}^{(5/3)}\right) & = \frac{1}{2} \left(\textrm{Tr}\, R^3 \right)^{2/3}\cdot \tau_{RR}\, ,\\
  \textrm{Im}\left( F_{S^1 \times S^2}^{(5/3)}\right) & \,\propto\,\, \frac{1}{2} \left(\textrm{Tr}\, R^3 \right)^{2/3}\cdot \kappa_{RR}\, .
\end{split}
\end{align}
Nevertheless,  we have obtained the large-$N$ expression of the free energy in \eqref{eq:A-twist logZ 5/3 large-N}:
\begin{displaymath}
  \textrm{log}\, Z_{S^1 \times S^2}^{\text{(5/3)}} = - \frac{3^{7/6}}{10} \left(1 - \frac{i}{\sqrt{3}} \right)\, \left(\frac{k}{\sum_{I=1}^{N_\chi} g_+ (\Delta_I)} \right)^{\frac{1}{3}}\, N^{\frac{5}{3}}\, \left(\sum_{I=1}^{N_\chi} (\mathfrak{n}_I - 1)\, g'_+ (\Delta_I) + |\mathcal{G}|\, \frac{\pi^2}{3}\right)\, ,
\end{displaymath}
which can also be written in a form similar to \eqref{eq:logZ IIA central charges}:
\be\label{eq:logZ 5/3 large-N}
  \textrm{log}\, Z_{S^1 \times S^2}^{\text{(5/3)}} (\mathfrak{n}_I,\, \Delta_I) = - \frac{3^{7/6}\, \pi}{5 \times 2^{10/3}} \left(1 - \frac{i}{\sqrt{3}} \right)\, (k N)^{1/3}\, \frac{c_r (\mathfrak{n}_I,\, \Delta_I)}{a (\Delta_I)^{1/3}}\, .
\ee
Based on \eqref{eq:F 5/3 expected} and \eqref{eq:logZ 5/3 large-N},  we expect that the same relation \eqref{eq:relation between c,  a and tau} should hold for the family of theories with the scaling $\sim N^{5/3}$:
\begin{displaymath}
  \frac{1}{2} \left(\textrm{Tr}\, R^3 \right)^{2/3}\cdot \tau_{RR}^{\textrm{min}} = - \frac{3^{7/6}\, \pi}{5 \times 2^{10/3}}\, (k N)^{1/3}\, \left[\frac{c_r (\frac{1}{\pi}\, \Delta_I,\, \Delta_I)}{a (\Delta_I)^{1/3}} \right]_{\textrm{min}}\, ,
\end{displaymath}
where $a$ and $c_r$ are still the 4d anomaly coefficient of the ``parent'' $\mathcal{N}=1$ superconformal field theory on $S^2 \times T^2$ and the 2d right-moving central charge $c_r$ of the $\mathcal{N}=(0,2)$ theory on $T^2$ respectively.  This reconfirms our speculation at the end of Sec.~\ref{sec:2D},  which becomes manifest for the 4d ``parent'' SCFTs from D3-branes at toric singularities.  As shown in \cite{Amariti:2021cpk} from the gauged supergravity,  the relation
\be
  \tau_{RR}^{\textrm{min}} = \frac{4}{\pi^2}\, F_{S^3}^{\textrm{max}}
\ee
holds for the massive IIA supergravity compactified on $S^6$,  we expect that this relation should also hold for the general class of theories with the scaling $\sim N^{5/3}$.

\subsection{A 3d Cardy-Like Formula}

Based on the examples discussed in the previous subsection,  we conjecture that the 3d $\mathcal{N}=2$ SCFT defined on $S_\beta^1 \times S^2$ to have a universal behavior in the Cardy-like limit \eqref{eq:CardyLimit},  which we call the 3d Cardy-like formula.  Namely,  the renormalized free energy of the 3d large-$N$ $\mathcal{N}=2$ SCFT has the following universal leading-order expression in the Cardy-like limit:
\be
  \textrm{Re}\left( F_{S^1 \times S^2} \right) = - \frac{1}{2} (m \overline{L})^2\, \tau_{aa} + \mathcal{O} (\beta)\, ,
\ee
where $m$ and $\overline{L}$ denote the background value of the scalar in the vector multiplet and the characteristic length scale of the 3d curved background respectively.  For instance,  $\overline{L}$ is $\beta$ for the non-rotating $S^2$,  while $\bar{L} = \beta - 2 \pi i/ \Omega$ for the rotating $S^2$.  Moreover,  $\tau_{aa}$ is the coefficient of the current-current two-point correlation function,  which can be either $\tau_{ff}$ or $\tau_{RR}$,  depending on whether the 3d SCFT has extra flavor symmetries.

Compared to $\textrm{Re}\left( F_{S^1 \times S^2} \right)$,  the imaginary part $\textrm{Im}\left( F_{S^1 \times S^2} \right)$ can be subleading in $\beta$,  but it also obeys a universal behavior:
\be
  \textrm{Im}\left( F_{S^1 \times S^2} \right) \,\,\propto\,\, - \frac{1}{2} (m \overline{L})^2\, \kappa_{aa}\, ,
\ee
where $\kappa_{aa}$ is the coefficient of the current-current two-point correlation function from Chern-Simons contact terms,  which can be either $\kappa_{ff}$ or $\kappa_{RR}$,  depending on whether the 3d SCFT has additional flavor symmetries.

\section{Discussion}\label{sec:Discussion}

In this paper, we consider 3d $\mathcal{N}=2$ SCFTs on $S_\beta^1 \times S^2$ and study their universal behavior in a Cardy-like limit $\beta \to 0$ (for the non-rotating case) or $\omega \sim \beta \to 0$ (for the rotating case).  We have exploited the viewpoint of the 3d rigid supersymmetric backgrounds for $S_\beta^1 \times S^2$,  which is amenable to a 2d EFT interpretation, to emphasize the latter in the Cardy-like limit.  We found a universal leading-in-$N$ behavior for the free energy with conceptually different origins from the 2d and 3d points of view; the 2d point of view allows for a direct evaluation via Bethe Ansatz solutions while for the 3d approach, we used properties of current correlators.  All the results in this paper also hold for the 3d SCFTs with $\mathcal{N} > 2$.  In this paper,  we have only focused on the 3d SCFTs on $S_\beta^1 \times S^2$,  which can be readily generalized to $S_\beta^1 \times \Sigma_g$ \cite{Closset:2019hyt}.  We anticipate that a map similar to \eqref{eq:Map} should exist between the 3d supersymmetric partition functions on $S_\beta^1 \times \Sigma_g$ with appropriate background fields and the topologically twisted indices on $S_\beta^1 \times \Sigma_g$,  as exploited for black holes in \cite{Azzurli:2017kxo}.

Related to the AdS black hole's microstate counting, we have found a map between a particular version of the 3d supersymmetric partition function on the $S^1_\beta \times ({\rm rotating}\,  S^2)$ background and the topologically twisted index in this paper.  The specific choice of the rigid supersymmetric  background connects,  through a series of identifications,  to a superconformal-index-like partition function.  This observation unveils a direct connection among different quantities computed in the literature for 3d SCFTs (e.g.,  superconformal index,  topologically twisted index, supersymmetric partition function on the generalized A-twisted background,  etc.).  Our rigid supergravity explanation sheds light on the ubiquity of certain structures at leading order when evaluating partition functions related to the entropy of AdS black holes.  It is worth highlighting that in some cases \cite{Hosseini:2016tor, Hosseini:2016ume, PandoZayas:2020iqr},  the relations among indices can go beyond the large-$N$ limit.  We believe it would be interesting to pursue interrelations among various supersymmetric partition functions more systematically, including at finite $N$.  Having associated the  3d Cardy-like formula with SCFT quantities,  namely,  with the coefficient $\tau_{aa}$ of the current-current correlators,  is a solid first step in this direction.

This paper also provides support and concrete examples to the conceptual discussions in \cite{Aharony:2017adm}.  As shown in \cite{Aharony:2017adm},  one obtains multiple 2d theories upon compactification from 3d,  depending on how the mass parameters scale.  When computing a specific quantity,  the authors of \cite{Aharony:2017adm} argue that one of the 2d theories usually provides the dominant contribution.  In this paper,  it has been possible to extract the dominant contribution in a Cardy-like limit to the supersymmetric partition function on the generalized A-twisted background.

There are some other classes of theories that have not been discussed in this paper.  For instance,  there is the 3d  $\mathcal{N}=4$ $T [SU(N)]$ SCFT \cite{Nishioka:2011dq,  Assel:2011xz},  whose large-$N$ free energy scales as $N^2\, \textrm{log}\, N$ \cite{Assel:2012cp,  Assel:2013lpa,  Coccia:2020cku}.  In principle,  the methods in this paper (supersymmetric partition function on a generalized A-twisted background,  3d Cardy-like formula,  etc.) can be applied to this theory with some additional care (e.g.,  the on-shell twisted superpotential),  and we expect similar results in the literature can be reproduced.

Given that the main interest in the partition functions we discussed in this paper stems from their relevance for computing the microscopic entropy of the dual asymptotically AdS$_4$ black holes, it is natural to wonder about the gravity implications of our 3d and 2d treatments. It has been argued that certain Cardy-like limits in field theory can be interpreted on the gravity side as near-horizon limits. This analysis was first performed for a certain limit of ${\cal N}=4$ SYM  in \cite{Nian:2020qsk} and subsequently discussed for all extremal asymptotically AdS black holes in dimension four, five, six, and seven \cite{David:2020ems}, and near-extremal AdS$_4$ in \cite{David:2020jhp}.  Moreover, some generalized Cardy limits have recently been explored from the field theory and gravity sides \cite{BenettiGenolini:2023rkq}.  There is an established tradition of describing near-horizon geometries in terms of putative dual CFT$_2$ theories dating back to \cite{Larsen:1997ge}. In this paper, we have shown that the 2d theory, obtained as a limit of a 3d theory,  is sufficient to provide the microscopic entropy of the dual black holes.  It would be quite interesting to explore if the 2d theory becomes conformal in a certain limit and,  more importantly,  if it can provide a description of the dynamical aspects of the black hole.

Our results in this paper complement an interesting development presented in \cite{Benini:2022bwa},   where a 3d ${\cal N}=2$ SCFT was reduced on $S^2$, and the resulting 1d quantum mechanics was shown to reproduce, in a particular approximation, the topologically twisted index of the 3d theory. Presumably, this 1d QM can capture some of the dynamics dual to the AdS$_2$ throat region on the gravity side. It would be quite interesting to investigate this picture further,  particularly in conjunction with the 2d picture we provided in this manuscript. We hope to address these exciting questions in the future.

\section*{\bf Acknowledgments}
We would like to thank Arash Arabi Ardehali,  Chi-Ming Chang,  Sunjin Choi,  Alfredo Gonz\'alez Lezcano,  Junho Hong,  Matthew Heydeman,  Gustavo Joaqu\'in Turiaci,  and Nathan Seiberg for the discussions.  The work of A.A. and A.S. has been supported in part by the Italian Ministero dell'Istruzione,  Universit\`a e Ricerca (MIUR), in part by Istituto Nazionale di Fisica Nucleare (INFN) through the “Gauge Theories, Strings, Supergravity” (GSS) research project and in part by MIUR-PRIN contract 2017CC72MK-003.  J.N.  was supported in part by the NSFC under grants No.~12375067 and No.~12147103.  L.P.Z.  is partially supported by the U.S. Department of Energy under grant DE-SC0007859; he also acknowledges support from an IBM Einstein Fellowship at the Institute for Advanced Study during 2022/2023.  A.A. and A.S. are thankful to ICTP for warm hospitality in Trieste, in the framework of {\it giornate-persona ICTP-INFN}, during the initial stages of this collaboration.

\bibliographystyle{utphys}
\bibliography{2dEFT}

\providecommand{\href}[2]{#2}\begingroup\raggedright\begin{thebibliography}{10}

\bibitem{Cabo-Bizet:2018ehj}
A.~Cabo-Bizet, D.~Cassani, D.~Martelli, and S.~Murthy, ``{Microscopic origin of the Bekenstein-Hawking entropy of supersymmetric AdS$_{5}$ black holes},'' \href{http://dx.doi.org/10.1007/JHEP10(2019)062}{{\em JHEP} {\bfseries 10} (2019) 062}, \href{http://arxiv.org/abs/1810.11442}{{\ttfamily arXiv:1810.11442 [hep-th]}}.

\bibitem{Choi:2018hmj}
S.~Choi, J.~Kim, S.~Kim, and J.~Nahmgoong, ``{Large AdS black holes from QFT},''
\href{http://arxiv.org/abs/1810.12067}{{\ttfamily arXiv:1810.12067 [hep-th]}}.

\bibitem{Benini:2018ywd}
F.~Benini and P.~Milan, ``{Black Holes in 4D $\mathcal{N}$=4 Super-Yang-Mills Field Theory},'' \href{http://dx.doi.org/10.1103/PhysRevX.10.021037}{{\em Phys. Rev. X} {\bfseries 10} no.~2, (2020) 021037}, \href{http://arxiv.org/abs/1812.09613}{{\ttfamily arXiv:1812.09613 [hep-th]}}.

\bibitem{Choi:2019miv}
S.~Choi and S.~Kim, ``{Large AdS$_{6}$ black holes from CFT$_{5}$},'' \href{http://dx.doi.org/10.1007/JHEP08(2024)228}{{\em JHEP} {\bfseries 08} (2024) 228}, \href{http://arxiv.org/abs/1904.01164}{{\ttfamily arXiv:1904.01164 [hep-th]}}.

\bibitem{Kantor:2019lfo}
G.~K\'antor, C.~Papageorgakis, and P.~Richmond, ``{AdS$_{7}$ black-hole entropy and 5D $ \mathcal{N} $ = 2 Yang-Mills},'' \href{http://dx.doi.org/10.1007/JHEP01(2020)017}{{\em JHEP} {\bfseries 01} (2020) 017}, \href{http://arxiv.org/abs/1907.02923}{{\ttfamily arXiv:1907.02923 [hep-th]}}.

\bibitem{Nahmgoong:2019hko}
J.~Nahmgoong, ``{6d superconformal Cardy formulas},'' \href{http://dx.doi.org/10.1007/JHEP02(2021)092}{{\em JHEP} {\bfseries 02} (2021) 092}, \href{http://arxiv.org/abs/1907.12582}{{\ttfamily arXiv:1907.12582 [hep-th]}}.

\bibitem{Choi:2019zpz}
S.~Choi, C.~Hwang, and S.~Kim, ``{Quantum vortices, M2-branes and black holes},'' \href{http://dx.doi.org/10.1007/JHEP09(2024)096}{{\em JHEP} {\bfseries 09} (2024) 096}, \href{http://arxiv.org/abs/1908.02470}{{\ttfamily arXiv:1908.02470 [hep-th]}}.

\bibitem{Nian:2019pxj}
J.~Nian and L.~A. Pando~Zayas, ``{Microscopic entropy of rotating electrically charged AdS$_{4}$ black holes from field theory localization},'' \href{http://dx.doi.org/10.1007/JHEP03(2020)081}{{\em JHEP} {\bfseries 03} (2020) 081}, \href{http://arxiv.org/abs/1909.07943}{{\ttfamily arXiv:1909.07943 [hep-th]}}.

\bibitem{Cardy:1986ie}
J.~L. Cardy, ``{Operator Content of Two-Dimensional Conformally Invariant Theories},'' \href{http://dx.doi.org/10.1016/0550-3213(86)90552-3}{{\em Nucl. Phys. B} {\bfseries 270} (1986) 186--204}.

\bibitem{DiPietro:2014bca}
L.~Di~Pietro and Z.~Komargodski, ``{Cardy formulae for SUSY theories in $d =$ 4 and $d =$ 6},'' \href{http://dx.doi.org/10.1007/JHEP12(2014)031}{{\em JHEP} {\bfseries 12} (2014) 031}, \href{http://arxiv.org/abs/1407.6061}{{\ttfamily arXiv:1407.6061 [hep-th]}}.

\bibitem{ArabiArdehali:2015iow}
A.~Arabi~Ardehali, J.~T. Liu, and P.~Szepietowski, ``{High-Temperature Expansion of Supersymmetric Partition Functions},'' \href{http://dx.doi.org/10.1007/JHEP07(2015)113}{{\em JHEP} {\bfseries 07} (2015) 113}, \href{http://arxiv.org/abs/1502.07737}{{\ttfamily arXiv:1502.07737 [hep-th]}}.

\bibitem{Chang:2019uag}
C.-M. Chang, M.~Fluder, Y.-H. Lin, and Y.~Wang, ``{Proving the 6d Cardy Formula and Matching Global Gravitational Anomalies},'' \href{http://dx.doi.org/10.21468/SciPostPhys.11.2.036}{{\em SciPost Phys.} {\bfseries 11} no.~2, (2021) 036}, \href{http://arxiv.org/abs/1910.10151}{{\ttfamily arXiv:1910.10151 [hep-th]}}.

\bibitem{Honda:2019cio}
M.~Honda, ``{Quantum Black Hole Entropy from 4d Supersymmetric Cardy formula},'' \href{http://dx.doi.org/10.1103/PhysRevD.100.026008}{{\em Phys. Rev. D} {\bfseries 100} no.~2, (2019) 026008}, \href{http://arxiv.org/abs/1901.08091}{{\ttfamily arXiv:1901.08091 [hep-th]}}.

\bibitem{ArabiArdehali:2019tdm}
A.~Arabi~Ardehali, ``{Cardy-like asymptotics of the 4d $ \mathcal{N}=4 $ index and AdS$_{5}$ blackholes},'' \href{http://dx.doi.org/10.1007/JHEP06(2019)134}{{\em JHEP} {\bfseries 06} (2019) 134}, \href{http://arxiv.org/abs/1902.06619}{{\ttfamily arXiv:1902.06619 [hep-th]}}.

\bibitem{Kim:2019yrz}
J.~Kim, S.~Kim, and J.~Song, ``{A 4d $ \mathcal{N} $ = 1 Cardy Formula},'' \href{http://dx.doi.org/10.1007/JHEP01(2021)025}{{\em JHEP} {\bfseries 01} (2021) 025}, \href{http://arxiv.org/abs/1904.03455}{{\ttfamily arXiv:1904.03455 [hep-th]}}.

\bibitem{Cassani:2021fyv}
D.~Cassani and Z.~Komargodski, ``{EFT and the SUSY Index on the 2nd Sheet},'' \href{http://dx.doi.org/10.21468/SciPostPhys.11.1.004}{{\em SciPost Phys.} {\bfseries 11} (2021) 004}, \href{http://arxiv.org/abs/2104.01464}{{\ttfamily arXiv:2104.01464 [hep-th]}}.

\bibitem{ArabiArdehali:2021nsx}
A.~Arabi~Ardehali and S.~Murthy, ``{The 4d superconformal index near roots of unity and 3d Chern-Simons theory},'' \href{http://dx.doi.org/10.1007/JHEP10(2021)207}{{\em JHEP} {\bfseries 10} (2021) 207}, \href{http://arxiv.org/abs/2104.02051}{{\ttfamily arXiv:2104.02051 [hep-th]}}.

\bibitem{GonzalezLezcano:2020yeb}
A.~Gonz\'alez~Lezcano, J.~Hong, J.~T. Liu, and L.~A. Pando~Zayas, ``{Sub-leading Structures in Superconformal Indices: Subdominant Saddles and Logarithmic Contributions},'' \href{http://dx.doi.org/10.1007/JHEP01(2021)001}{{\em JHEP} {\bfseries 01} (2021) 001}, \href{http://arxiv.org/abs/2007.12604}{{\ttfamily arXiv:2007.12604 [hep-th]}}.

\bibitem{Amariti:2020jyx}
A.~Amariti, M.~Fazzi, and A.~Segati, ``{The SCI of $ \mathcal{N} $ = 4 USp(2N$_{c}$) and SO(N$_{c}$) SYM as a matrix integral},'' \href{http://dx.doi.org/10.1007/JHEP06(2021)132}{{\em JHEP} {\bfseries 06} (2021) 132}, \href{http://arxiv.org/abs/2012.15208}{{\ttfamily arXiv:2012.15208 [hep-th]}}.

\bibitem{Amariti:2021ubd}
A.~Amariti, M.~Fazzi, and A.~Segati, ``{Expanding on the Cardy-like limit of the SCI of 4d $ \mathcal{N} $ = 1 ABCD SCFTs},'' \href{http://dx.doi.org/10.1007/JHEP07(2021)141}{{\em JHEP} {\bfseries 07} (2021) 141}, \href{http://arxiv.org/abs/2103.15853}{{\ttfamily arXiv:2103.15853 [hep-th]}}.

\bibitem{Gang:2019uay}
D.~Gang, N.~Kim, and L.~A. Pando~Zayas, ``{Precision Microstate Counting for the Entropy of Wrapped M5-branes},'' \href{http://dx.doi.org/10.1007/JHEP03(2020)164}{{\em JHEP} {\bfseries 03} (2020) 164}, \href{http://arxiv.org/abs/1905.01559}{{\ttfamily arXiv:1905.01559 [hep-th]}}.

\bibitem{Bobev:2019zmz}
N.~Bobev and P.~M. Crichigno, ``{Universal spinning black holes and theories of class $ \mathcal{R} $},'' \href{http://dx.doi.org/10.1007/JHEP12(2019)054}{{\em JHEP} {\bfseries 12} (2019) 054}, \href{http://arxiv.org/abs/1909.05873}{{\ttfamily arXiv:1909.05873 [hep-th]}}.

\bibitem{Benini:2019dyp}
F.~Benini, D.~Gang, and L.~A. Pando~Zayas, ``{Rotating Black Hole Entropy from M5 Branes},'' \href{http://dx.doi.org/10.1007/JHEP03(2020)057}{{\em JHEP} {\bfseries 03} (2020) 057}, \href{http://arxiv.org/abs/1909.11612}{{\ttfamily arXiv:1909.11612 [hep-th]}}.

\bibitem{Choi:2019dfu}
S.~Choi and C.~Hwang, ``{Universal 3d Cardy Block and Black Hole Entropy},'' \href{http://dx.doi.org/10.1007/JHEP03(2020)068}{{\em JHEP} {\bfseries 03} (2020) 068}, \href{http://arxiv.org/abs/1911.01448}{{\ttfamily arXiv:1911.01448 [hep-th]}}.

\bibitem{Hosseini:2022vho}
S.~M. Hosseini and A.~Zaffaroni, ``{The large N limit of topologically twisted indices: a direct approach},'' \href{http://dx.doi.org/10.1007/JHEP12(2022)025}{{\em JHEP} {\bfseries 12} (2022) 025}, \href{http://arxiv.org/abs/2209.09274}{{\ttfamily arXiv:2209.09274 [hep-th]}}.

\bibitem{GonzalezLezcano:2022hcf}
A.~Gonz\'alez~Lezcano, M.~Jerdee, and L.~A. Pando~Zayas, ``{Cardy expansion of 3d superconformal indices and corrections to the dual black hole entropy},'' \href{http://dx.doi.org/10.1007/JHEP01(2023)044}{{\em JHEP} {\bfseries 01} (2023) 044}, \href{http://arxiv.org/abs/2210.12065}{{\ttfamily arXiv:2210.12065 [hep-th]}}.

\bibitem{Bobev:2022wem}
N.~Bobev, S.~Choi, J.~Hong, and V.~Reys, ``{Large N superconformal indices for 3d holographic SCFTs},'' \href{http://dx.doi.org/10.1007/JHEP02(2023)027}{{\em JHEP} {\bfseries 02} (2023) 027}, \href{http://arxiv.org/abs/2210.15326}{{\ttfamily arXiv:2210.15326 [hep-th]}}.

\bibitem{Bobev:2023lkx}
N.~Bobev, J.~Hong, and V.~Reys, ``{Large N partition functions of 3d holographic SCFTs},'' \href{http://dx.doi.org/10.1007/JHEP08(2023)119}{{\em JHEP} {\bfseries 08} (2023) 119}, \href{http://arxiv.org/abs/2304.01734}{{\ttfamily arXiv:2304.01734 [hep-th]}}.

\bibitem{Bobev:2024mqw}
N.~Bobev, S.~Choi, J.~Hong, and V.~Reys, ``{Superconformal indices of 3d $ \mathcal{N} $ = 2 SCFTs and holography},'' \href{http://dx.doi.org/10.1007/JHEP10(2024)121}{{\em JHEP} {\bfseries 10} (2024) 121}, \href{http://arxiv.org/abs/2407.13177}{{\ttfamily arXiv:2407.13177 [hep-th]}}.

\bibitem{Witten:1988ze}
E.~Witten, ``{Topological Quantum Field Theory},'' \href{http://dx.doi.org/10.1007/BF01223371}{{\em Commun. Math. Phys.} {\bfseries 117} (1988) 353}.

\bibitem{Closset:2017zgf}
C.~Closset, H.~Kim, and B.~Willett, ``{Supersymmetric partition functions and the three-dimensional A-twist},'' \href{http://dx.doi.org/10.1007/JHEP03(2017)074}{{\em JHEP} {\bfseries 03} (2017) 074}, \href{http://arxiv.org/abs/1701.03171}{{\ttfamily arXiv:1701.03171 [hep-th]}}.

\bibitem{Closset:2019hyt}
C.~Closset and H.~Kim, ``{Three-dimensional \ensuremath{\mathscr{N}} = 2 supersymmetric gauge theories and partition functions on Seifert manifolds: A review},'' \href{http://dx.doi.org/10.1142/S0217751X19300114}{{\em Int. J. Mod. Phys. A} {\bfseries 34} no.~23, (2019) 1930011}, \href{http://arxiv.org/abs/1908.08875}{{\ttfamily arXiv:1908.08875 [hep-th]}}.

\bibitem{Hwang:2018riu}
C.~Hwang, S.~Lee, and P.~Yi, ``{Holonomy Saddles and Supersymmetry},'' \href{http://dx.doi.org/10.1103/PhysRevD.97.125013}{{\em Phys. Rev. D} {\bfseries 97} no.~12, (2018) 125013}, \href{http://arxiv.org/abs/1801.05460}{{\ttfamily arXiv:1801.05460 [hep-th]}}.

\bibitem{Benini:2015eyy}
F.~Benini, K.~Hristov, and A.~Zaffaroni, ``{Black hole microstates in AdS$_{4}$ from supersymmetric localization},'' \href{http://dx.doi.org/10.1007/JHEP05(2016)054}{{\em JHEP} {\bfseries 05} (2016) 054}, \href{http://arxiv.org/abs/1511.04085}{{\ttfamily arXiv:1511.04085 [hep-th]}}.

\bibitem{Festuccia:2011ws}
G.~Festuccia and N.~Seiberg, ``{Rigid Supersymmetric Theories in Curved Superspace},'' \href{http://dx.doi.org/10.1007/JHEP06(2011)114}{{\em JHEP} {\bfseries 06} (2011) 114},
\href{http://arxiv.org/abs/1105.0689}{{\ttfamily arXiv:1105.0689 [hep-th]}}.

\bibitem{Barnes:2005bw}
E.~Barnes, E.~Gorbatov, K.~A. Intriligator, and J.~Wright, ``{Current correlators and AdS/CFT geometry},'' \href{http://dx.doi.org/10.1016/j.nuclphysb.2005.10.013}{{\em Nucl. Phys. B} {\bfseries 732} (2006) 89--117}, \href{http://arxiv.org/abs/hep-th/0507146}{{\ttfamily arXiv:hep-th/0507146}}.

\bibitem{Closset:2012vg}
C.~Closset, T.~T. Dumitrescu, G.~Festuccia, Z.~Komargodski, and N.~Seiberg, ``{Contact Terms, Unitarity, and F-Maximization in Three-Dimensional Superconformal Theories},'' \href{http://dx.doi.org/10.1007/JHEP10(2012)053}{{\em JHEP} {\bfseries 10} (2012) 053}, \href{http://arxiv.org/abs/1205.4142}{{\ttfamily arXiv:1205.4142 [hep-th]}}.

\bibitem{Hosseini:2016tor}
S.~M. Hosseini and A.~Zaffaroni, ``{Large $N$ matrix models for 3d ${\cal N}=2$ theories: twisted index, free energy and black holes},'' \href{http://dx.doi.org/10.1007/JHEP08(2016)064}{{\em JHEP} {\bfseries 08} (2016) 064}, \href{http://arxiv.org/abs/1604.03122}{{\ttfamily arXiv:1604.03122 [hep-th]}}.

\bibitem{Crichigno:2020ouj}
P.~M. Crichigno and D.~Jain, ``{The 5d Superconformal Index at Large $N$ and Black Holes},'' \href{http://dx.doi.org/10.1007/JHEP09(2020)124}{{\em JHEP} {\bfseries 09} (2020) 124}, \href{http://arxiv.org/abs/2005.00550}{{\ttfamily arXiv:2005.00550 [hep-th]}}.

\bibitem{Closset:2012ru}
C.~Closset, T.~T. Dumitrescu, G.~Festuccia, and Z.~Komargodski, ``{Supersymmetric Field Theories on Three-Manifolds},'' \href{http://dx.doi.org/10.1007/JHEP05(2013)017}{{\em JHEP} {\bfseries 05} (2013) 017},
\href{http://arxiv.org/abs/1212.3388}{{\ttfamily arXiv:1212.3388 [hep-th]}}.

\bibitem{Nian:2013qwa}
J.~Nian, ``{Localization of Supersymmetric Chern-Simons-Matter Theory on a Squashed $S^3$ with $SU(2)\times U(1)$ Isometry},'' \href{http://dx.doi.org/10.1007/JHEP07(2014)126}{{\em JHEP} {\bfseries 07} (2014) 126},
\href{http://arxiv.org/abs/1309.3266}{{\ttfamily arXiv:1309.3266 [hep-th]}}.

\bibitem{Aharony:2017adm}
O.~Aharony, S.~S. Razamat, and B.~Willett, ``{From 3d duality to 2d duality},'' \href{http://dx.doi.org/10.1007/JHEP11(2017)090}{{\em JHEP} {\bfseries 11} (2017) 090}, \href{http://arxiv.org/abs/1710.00926}{{\ttfamily arXiv:1710.00926 [hep-th]}}.

\bibitem{Aharony:2008ug}
O.~Aharony, O.~Bergman, D.~L. Jafferis, and J.~Maldacena, ``{N=6 superconformal Chern-Simons-matter theories, M2-branes and their gravity duals},'' \href{http://dx.doi.org/10.1088/1126-6708/2008/10/091}{{\em JHEP} {\bfseries 10} (2008) 091},
\href{http://arxiv.org/abs/0806.1218}{{\ttfamily arXiv:0806.1218 [hep-th]}}.

\bibitem{Bhattacharya:2008bja}
J.~Bhattacharya and S.~Minwalla, ``{Superconformal Indices for N = 6 Chern Simons Theories},'' \href{http://dx.doi.org/10.1088/1126-6708/2009/01/014}{{\em JHEP} {\bfseries 01} (2009) 014},
\href{http://arxiv.org/abs/0806.3251}{{\ttfamily arXiv:0806.3251 [hep-th]}}.

\bibitem{Kim:2009wb}
S.~Kim, ``{The Complete superconformal index for N=6 Chern-Simons theory},'' \href{http://dx.doi.org/10.1016/j.nuclphysb.2012.07.015, 10.1016/j.nuclphysb.2009.06.025}{{\em Nucl. Phys.} {\bfseries B821} (2009) 241--284}, \href{http://arxiv.org/abs/0903.4172}{{\ttfamily arXiv:0903.4172 [hep-th]}}.
[Erratum: Nucl. Phys.B864,884(2012)].

\bibitem{Amariti:2023}
A.~Amariti, J.~Nian, L.~Pando~Zayas, A.~Segati, and H.~Zhang, ``in preparation,''.

\bibitem{Warner:1983vz}
N.~P. Warner, ``{Some New Extrema of the Scalar Potential of Gauged $N=8$ Supergravity},'' \href{http://dx.doi.org/10.1016/0370-2693(83)90383-0}{{\em Phys. Lett. B} {\bfseries 128} (1983) 169--173}.

\bibitem{Warner:1983du}
N.~P. Warner, ``{Some Properties of the Scalar Potential in Gauged Supergravity Theories},'' \href{http://dx.doi.org/10.1016/0550-3213(84)90286-4}{{\em Nucl. Phys. B} {\bfseries 231} (1984) 250--268}.

\bibitem{Bobev:2018uxk}
N.~Bobev, V.~S. Min, and K.~Pilch, ``{Mass-deformed ABJM and black holes in AdS$_{4}$},'' \href{http://dx.doi.org/10.1007/JHEP03(2018)050}{{\em JHEP} {\bfseries 03} (2018) 050}, \href{http://arxiv.org/abs/1801.03135}{{\ttfamily arXiv:1801.03135 [hep-th]}}.

\bibitem{Romans:1985tz}
L.~J. Romans, ``{Massive N=2a Supergravity in Ten-Dimensions},'' \href{http://dx.doi.org/10.1016/0370-2693(86)90375-8}{{\em Phys. Lett. B} {\bfseries 169} (1986) 374}.

\bibitem{Guarino:2015jca}
A.~Guarino, D.~L. Jafferis, and O.~Varela, ``{String Theory Origin of Dyonic N=8 Supergravity and Its Chern-Simons Duals},'' \href{http://dx.doi.org/10.1103/PhysRevLett.115.091601}{{\em Phys. Rev. Lett.} {\bfseries 115} no.~9, (2015) 091601}, \href{http://arxiv.org/abs/1504.08009}{{\ttfamily arXiv:1504.08009 [hep-th]}}.

\bibitem{Fluder:2015eoa}
M.~Fluder and J.~Sparks, ``{D2-brane Chern-Simons theories: F-maximization = a-maximization},'' \href{http://dx.doi.org/10.1007/JHEP01(2016)048}{{\em JHEP} {\bfseries 01} (2016) 048}, \href{http://arxiv.org/abs/1507.05817}{{\ttfamily arXiv:1507.05817 [hep-th]}}.

\bibitem{Hosseini:2017fjo}
S.~M. Hosseini, K.~Hristov, and A.~Passias, ``{Holographic microstate counting for AdS$_{4}$ black holes in massive IIA supergravity},'' \href{http://dx.doi.org/10.1007/JHEP10(2017)190}{{\em JHEP} {\bfseries 10} (2017) 190}, \href{http://arxiv.org/abs/1707.06884}{{\ttfamily arXiv:1707.06884 [hep-th]}}.

\bibitem{Benini:2017oxt}
F.~Benini, H.~Khachatryan, and E.~Milan, ``{Black hole entropy in massive Type IIA},'' \href{http://dx.doi.org/10.1088/1361-6382/aa9f5b}{{\em Class. Quant. Grav.} {\bfseries 35} no.~3, (2018) 035004}, \href{http://arxiv.org/abs/1707.06886}{{\ttfamily arXiv:1707.06886 [hep-th]}}.

\bibitem{Liu:2018bac}
J.~T. Liu, L.~A. Pando~Zayas, and S.~Zhou, ``{Subleading Microstate Counting in the Dual to Massive Type IIA},'' \href{http://arxiv.org/abs/1808.10445}{{\ttfamily arXiv:1808.10445 [hep-th]}}.

\bibitem{Hosseini:2016cyf}
S.~M. Hosseini, A.~Nedelin, and A.~Zaffaroni, ``{The Cardy limit of the topologically twisted index and black strings in AdS$_{5}$},'' \href{http://dx.doi.org/10.1007/JHEP04(2017)014}{{\em JHEP} {\bfseries 04} (2017) 014}, \href{http://arxiv.org/abs/1611.09374}{{\ttfamily arXiv:1611.09374 [hep-th]}}.

\bibitem{Butti:2005vn}
A.~Butti and A.~Zaffaroni, ``{R-charges from toric diagrams and the equivalence of a-maximization and Z-minimization},'' \href{http://dx.doi.org/10.1088/1126-6708/2005/11/019}{{\em JHEP} {\bfseries 11} (2005) 019}, \href{http://arxiv.org/abs/hep-th/0506232}{{\ttfamily arXiv:hep-th/0506232}}.

\bibitem{Benvenuti:2006xg}
S.~Benvenuti, L.~A. Pando~Zayas, and Y.~Tachikawa, ``{Triangle anomalies from Einstein manifolds},'' \href{http://dx.doi.org/10.4310/ATMP.2006.v10.n3.a4}{{\em Adv. Theor. Math. Phys.} {\bfseries 10} no.~3, (2006) 395--432}, \href{http://arxiv.org/abs/hep-th/0601054}{{\ttfamily arXiv:hep-th/0601054}}.

\bibitem{David:2020ems}
M.~David, J.~Nian, and L.~A. Pando~Zayas, ``{Gravitational Cardy Limit and AdS Black Hole Entropy},'' \href{http://dx.doi.org/10.1007/JHEP11(2020)041}{{\em JHEP} {\bfseries 11} (2020) 041}, \href{http://arxiv.org/abs/2005.10251}{{\ttfamily arXiv:2005.10251 [hep-th]}}.

\bibitem{Amariti:2015ybz}
A.~Amariti and A.~Gnecchi, ``{3D $\tau_{RR}$-minimization in AdS$_4$ gauged supergravity},'' \href{http://dx.doi.org/10.1007/JHEP07(2016)006}{{\em JHEP} {\bfseries 07} (2016) 006}, \href{http://arxiv.org/abs/1511.08214}{{\ttfamily arXiv:1511.08214 [hep-th]}}.

\bibitem{Amariti:2021cpk}
A.~Amariti and A.~Gnecchi, ``{\ensuremath{\tau}$_{RR}$ minimization in presence of hypermultiplets},'' \href{http://dx.doi.org/10.1007/JHEP03(2022)166}{{\em JHEP} {\bfseries 03} (2022) 166}, \href{http://arxiv.org/abs/2107.01195}{{\ttfamily arXiv:2107.01195 [hep-th]}}.

\bibitem{Azzurli:2017kxo}
F.~Azzurli, N.~Bobev, P.~M. Crichigno, V.~S. Min, and A.~Zaffaroni, ``{A universal counting of black hole microstates in AdS$_{4}$},'' \href{http://dx.doi.org/10.1007/JHEP02(2018)054}{{\em JHEP} {\bfseries 02} (2018) 054}, \href{http://arxiv.org/abs/1707.04257}{{\ttfamily arXiv:1707.04257 [hep-th]}}.

\bibitem{Hosseini:2016ume}
S.~M. Hosseini and N.~Mekareeya, ``{Large $N$ topologically twisted index: necklace quivers, dualities, and Sasaki-Einstein spaces},'' \href{http://dx.doi.org/10.1007/JHEP08(2016)089}{{\em JHEP} {\bfseries 08} (2016) 089}, \href{http://arxiv.org/abs/1604.03397}{{\ttfamily arXiv:1604.03397 [hep-th]}}.

\bibitem{PandoZayas:2020iqr}
L.~A. Pando~Zayas and Y.~Xin, ``{Universal logarithmic behavior in microstate counting and the dual one-loop entropy of $AdS_4$ black holes},'' \href{http://dx.doi.org/10.1103/PhysRevD.103.026003}{{\em Phys. Rev. D} {\bfseries 103} no.~2, (2021) 026003}, \href{http://arxiv.org/abs/2008.03239}{{\ttfamily arXiv:2008.03239 [hep-th]}}.

\bibitem{Nishioka:2011dq}
T.~Nishioka, Y.~Tachikawa, and M.~Yamazaki, ``{3d Partition Function as Overlap of Wavefunctions},'' \href{http://dx.doi.org/10.1007/JHEP08(2011)003}{{\em JHEP} {\bfseries 08} (2011) 003}, \href{http://arxiv.org/abs/1105.4390}{{\ttfamily arXiv:1105.4390 [hep-th]}}.

\bibitem{Assel:2011xz}
B.~Assel, C.~Bachas, J.~Estes, and J.~Gomis, ``{Holographic Duals of D=3 N=4 Superconformal Field Theories},'' \href{http://dx.doi.org/10.1007/JHEP08(2011)087}{{\em JHEP} {\bfseries 08} (2011) 087}, \href{http://arxiv.org/abs/1106.4253}{{\ttfamily arXiv:1106.4253 [hep-th]}}.

\bibitem{Assel:2012cp}
B.~Assel, J.~Estes, and M.~Yamazaki, ``{Large N Free Energy of 3d N=4 SCFTs and $AdS_4/CFT_3$},'' \href{http://dx.doi.org/10.1007/JHEP09(2012)074}{{\em JHEP} {\bfseries 09} (2012) 074}, \href{http://arxiv.org/abs/1206.2920}{{\ttfamily arXiv:1206.2920 [hep-th]}}.

\bibitem{Assel:2013lpa}
B.~Assel, {\em {Holographic Duality for three-dimensional Super-conformal Field Theories}}.
\newblock PhD thesis, Ecole Normale Superieure, 2013.
\newblock \href{http://arxiv.org/abs/1307.4244}{{\ttfamily arXiv:1307.4244 [hep-th]}}.

\bibitem{Coccia:2020cku}
L.~Coccia, ``{Topologically twisted index of $T[SU(N)]$ at large $N$},'' \href{http://dx.doi.org/10.1007/JHEP05(2021)264}{{\em JHEP} {\bfseries 05} (2021) 264}, \href{http://arxiv.org/abs/2006.06578}{{\ttfamily arXiv:2006.06578 [hep-th]}}.

\bibitem{Nian:2020qsk}
J.~Nian and L.~A. Pando~Zayas, ``{Toward an effective CFT$_{2}$ from $ \mathcal{N} $ = 4 super Yang-Mills and aspects of Hawking radiation},'' \href{http://dx.doi.org/10.1007/JHEP07(2020)120}{{\em JHEP} {\bfseries 07} (2020) 120}, \href{http://arxiv.org/abs/2003.02770}{{\ttfamily arXiv:2003.02770 [hep-th]}}.

\bibitem{David:2020jhp}
M.~David and J.~Nian, ``{Universal entropy and hawking radiation of near-extremal AdS$_{4}$ black holes},'' \href{http://dx.doi.org/10.1007/JHEP04(2021)256}{{\em JHEP} {\bfseries 04} (2021) 256}, \href{http://arxiv.org/abs/2009.12370}{{\ttfamily arXiv:2009.12370 [hep-th]}}.

\bibitem{BenettiGenolini:2023rkq}
P.~Benetti~Genolini, A.~Cabo-Bizet, and S.~Murthy, ``{Supersymmetric phases of AdS$_{4}$/CFT$_{3}$},'' \href{http://dx.doi.org/10.1007/JHEP06(2023)125}{{\em JHEP} {\bfseries 06} (2023) 125}, \href{http://arxiv.org/abs/2301.00763}{{\ttfamily arXiv:2301.00763 [hep-th]}}.

\bibitem{Larsen:1997ge}
F.~Larsen, ``{A String model of black hole microstates},'' \href{http://dx.doi.org/10.1103/PhysRevD.56.1005}{{\em Phys. Rev. D} {\bfseries 56} (1997) 1005--1008}, \href{http://arxiv.org/abs/hep-th/9702153}{{\ttfamily arXiv:hep-th/9702153}}.

\bibitem{Benini:2022bwa}
F.~Benini, S.~Soltani, and Z.~Zhang, ``{A quantum mechanics for magnetic horizons},'' \href{http://dx.doi.org/10.1007/JHEP05(2023)070}{{\em JHEP} {\bfseries 05} (2023) 070}, \href{http://arxiv.org/abs/2212.00672}{{\ttfamily arXiv:2212.00672 [hep-th]}}.

\end{thebibliography}\endgroup
\end{document}